\documentclass[review]{elsarticle}

\usepackage{lineno,hyperref}
\modulolinenumbers[5]

\journal{Journal of \LaTeX\ Templates}

\begin{document}

\begin{frontmatter}

\title{Qualification tests of 997 8-inch photomultiplier tubes for the water Cherenkov detector array of the LHAASO experiment}


\author[add1,add2]{Kun Jiang}

\author[add1,add2]{Zebo Tang\corref{cor1}}
\cortext[cor1]{Corresponding author}
\ead{zbtang@ustc.edu.cn}

\author[add1,add2]{Xin Li\corref{cor1}}
\ead{li124ste@ustc.edu.cn}

\author[add1,add2]{Zehua Cao} 
\author[add1,add2]{Cheng Li}
\author[add1,add2]{Yang Li} 
\author[add1,add2]{Ziwei Li} 
\author[add1,add2]{Ziyang Li} 
\author[add1,add2]{Zheng Liang}
\author[add1,add2]{Pengzhong Lu} 
\author[add1,add2]{Kaifeng Shen}
\author[add1,add2]{Kaiyang Wang} 
\author[add1,add2]{Yan Wang} 
\author[add1,add2]{Xin Wu} 

\address[add1]{State Key Laboratory of Particle Detection and Electronics, University of Science and Technology of China, Hefei 230026, China}
\address[add2]{Department of Modern Physics, University of Science and Technology of China, Hefei 230026, China}

\begin{abstract}
The Large High-Altitude Air Shower Observatory (LHAASO) is being built at Haizi Mountain, Sichuan province of China at an altitude of 4410 meters. One of its main goals is to survey the northern sky for very-high-energy gamma ray sources via its ground-based water Cherenkov detector array (WCDA). 900 8-inch photomultiplier tubes (PMTs) CR365-02-1 from Beijing Hamamatsu Photon Techniques INC. (BHP) are installed in the WCDA, collecting Cherenkov photons produced by air shower particles crossing water. The design of the PMT base with a high dynamic range for CR365-02-1, the PMT batch test system, and the test results of 997 PMTs are presented in this paper. 
\end{abstract}


\end{frontmatter}





\section{Introduction}\label{intro}
The Large High Altitude Air Shower Observatory (LHAASO) project~\cite{Cao:2010zz} is a large-scale complex of astrophysics detectors being built at Haizi Mountain, Sichuan province of China at an altitude of 4410 meters. As shown in Fig.~\ref{fig:lhaaso}, the proposed detector arrays include~\cite{He2018Design}: 1 $km^2$ array (KM2A) composed of 5195 electromagnetic particle detectors (EDs) and 1171 muon detectors (MDs), 78,000 $m^2$ water Cherenkov detector array (WCDA), and 12 wide-field air Cherenkov/fluorescence telescopes (WFCTA). The scientific goals of LHAASO are (1) searching for galactic cosmic ray origins; (2) all sky survey for very-high-energy gamma ray sources; (3) energy spectrum and composition measurements of cosmic rays over a wide range covering the knee region~\cite{Cao:2014rla}.
{
\begin{figure}[htbp]
\begin{center}
\includegraphics[keepaspectratio,width=0.75\textwidth]{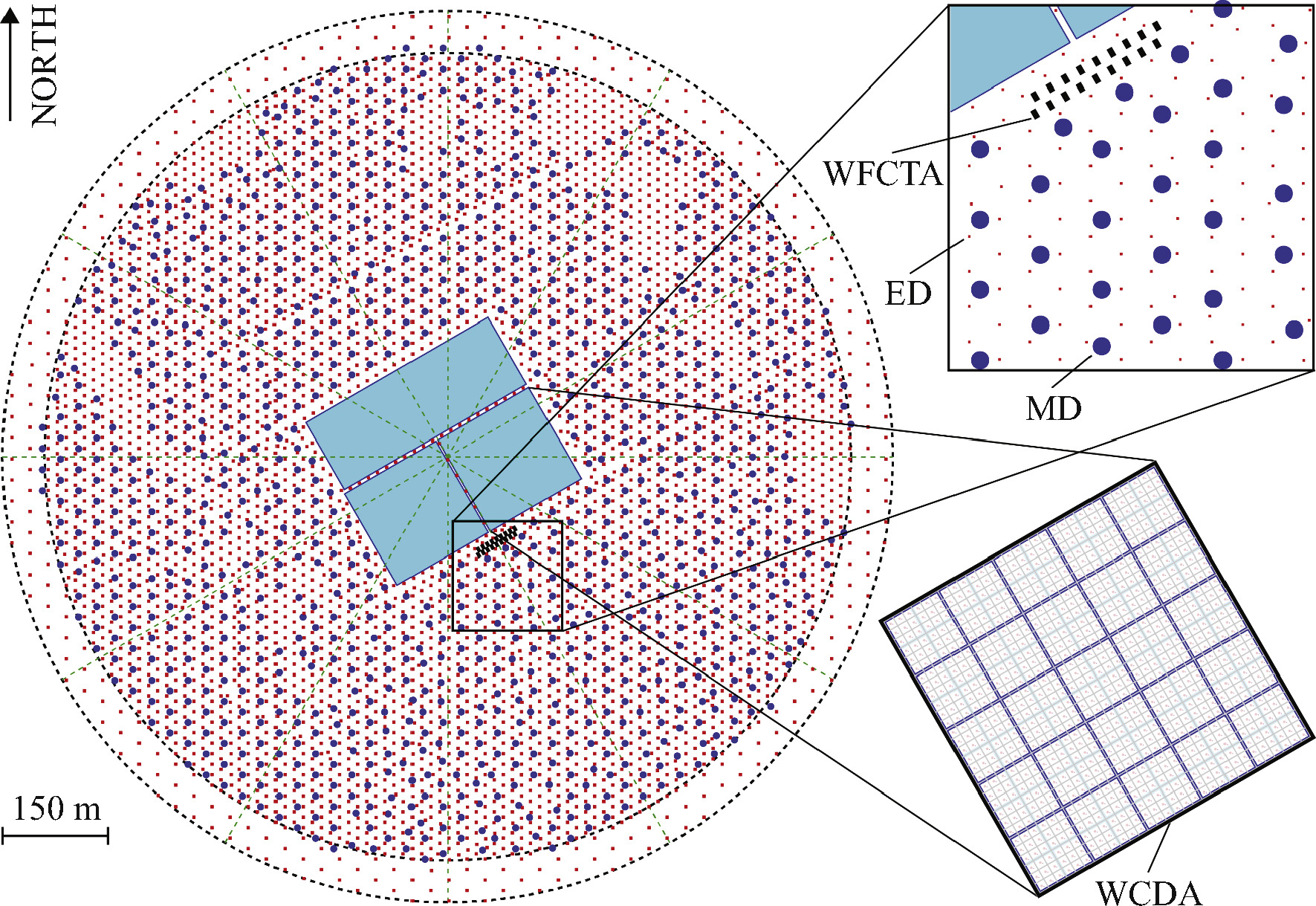}
\vspace*{+0mm}
\caption{The layout of LHAASO detectors.} \label{fig:lhaaso}
\end{center}
\end{figure}
}

As one of the major components of the LHAASO project, the WCDA targets gamma astronomy at the energy between 100 GeV and 30 TeV~\cite{An:2011za}. The WCDA covers an effective area of 78000 $m^2$ and is divided into 3 water ponds. Each water pond of the WCDA is partitioned by curtains into 5 m $\times$ 5 m detector cells with an effective water depth of 4 m. The first two water ponds with an effective area of 150 $\times$ 150 $m^2$ contain 900 detector cells each. 
The third water pond with an area of 300 $\times$ 110 $m^2$  contains 1320 detector cells. There are 3120 detector cells in total. 
A pair of 1.5- and 8-inch PMTs are installed at the bottom-center of each cell of the first water pond~\cite{Zhang:2020jgr}, collecting the Cherenkov light produced by air shower particles crossing water. For the other two water ponds, a pair of 3- and 20-inch PMTs are installed in each detector cell.

{
\begin{table*}
\centering
  \begin{tabular}{c|c}
    Parameters  &  Specifications  \\\hline
    Working gain &  $3\times10^6$\\
    Distribution of working voltage & Mean $\pm$  100 V, $<$ 2000 V \\ 
    Distribution of amplification voltage coefficient ($\beta$) &   Mean $\pm$ 0.5  \\
    Peak-to-valley ratio   &  $>$ 2.0  \\
    Quantum efficiency (405 nm)  &  $>$ 22\%  \\
    Transit time spread (FWHM)  &  $<$ 4 ns  \\
    Dark count rate ($>1/3$ PE)   &   $<$ 5000 cps \\
    Afterpulse rate ($100-10000~\textrm{ns}$)   &   $<$ 5\% \\
    Anode linearity (5\%)    &   $>$ 1000 PEs\\
    8th dynode linearity (5\%)   &   $>$ 4000 PEs\\
    Distribution of anode-to-dynode charge ratio  &Mean $\pm$ 15\%\\
\end{tabular}
	\caption{Specification requirements of CR365-02-1.}
  \label{tab:cut}
\end{table*}
}
According to the Monte Carlo simulation of gamma rays from the Crab Nebula, PMTs are required to have not only good single photoelectron (SPE) resolution but also good linearity up to 4000 photoelectrons (PEs)~\cite{wcdaicrc2011,An:2011za}. Fast timing characteristics are also important to achieve good reconstruction accuracy of the extensive air shower front. The specifications for 8-inch PMTs are listed in Table~\ref{tab:cut}. The PMTs are operated with a gain of $3\times10^6$. In the WCDA experiment, three PMTs with similar working voltages share one power supply, and are placed in adjacent cells. In order to remove the outliers and make it easier to group the PMTs, specifications of mean $\pm$ 100 V for working voltage and mean $\pm$ 0.5 for the amplification voltage coefficient are required. When a large number of shower particles pass through the water Cherenkov detector and the anode reaches its maximum dynamic range, the signal can be read from the dynode to extend the dynamic range. There is a readout overlapping range where signals can be read from both the anode and the dynode. The overlapping range is partially determined by the anode-to-dynode charge ratio. A specification of mean $\pm$ 15\% for the anode-to-dynode charge ratio can make sure that all PMTs have a uniform overlapping range.The 8-inch PMT selected by the WCDA is the CR365-02-1 made by Beijing Hamamatsu Photon Techniques INC. (BHP). In order to meet all of the requirements, we designed a special high dynamic range base circuit using two outputs from the CR365-02-1: one from the anode and the other one from the 8th dynode (Dy8). The ratio of the gain of the two outputs is tuned to about 50 to balance the dynamic range and overlapping range. The schematic of the base circuit is shown in Fig.~\ref{fig:base}(a). 
{
\begin{figure}[htbp]
\begin{center}
\includegraphics[keepaspectratio,width=0.98\textwidth]{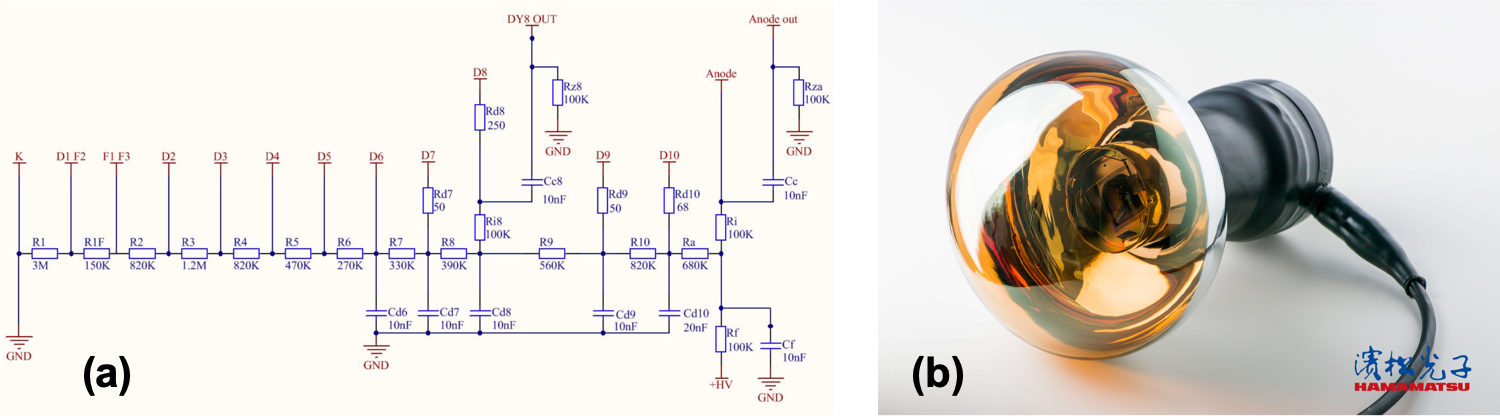}
\vspace*{+0mm}
\caption{(a) The base circuit used for the LHAASO-WCDA. (b) The PMT CR365-02-1 from BHP.} \label{fig:base}
\end{center}
\end{figure}
}
It is similar to what we used to evaluate another type of PMT for the LHAASO-WCDA~\cite{Zhao:2016utp}. 
The damping resistor Rd8 (250 $\Omega$) connected to the Dy8 has been fine tuned to minimize the oscillation of the pulse. 
The base circuit is produced and assembled to the PMT by BHP. The assembly of the CR365-02-1 is shown in Fig.~\ref{fig:base}(b). 
There are 3 30-meter-long cables, one for high voltage supply and two for signal outputs, bundled together and covered by waterproof material. 

\section{The test system}\label{testfacility}
The test system is built at University of Science and Technology of China in Hefei.
The system is designed to test 16 PMTs simultaneously during one test run.
Two calibrated PMTs stay in the setup to serve as references and monitor the stability of the whole system. 
The test procedure is pretty much automatic. 
East test run takes about 15 hours so the system can test 14 new PMTs per day. 
{
\begin{figure}[htbp]
\begin{center}
\includegraphics[keepaspectratio,width=0.90\textwidth]{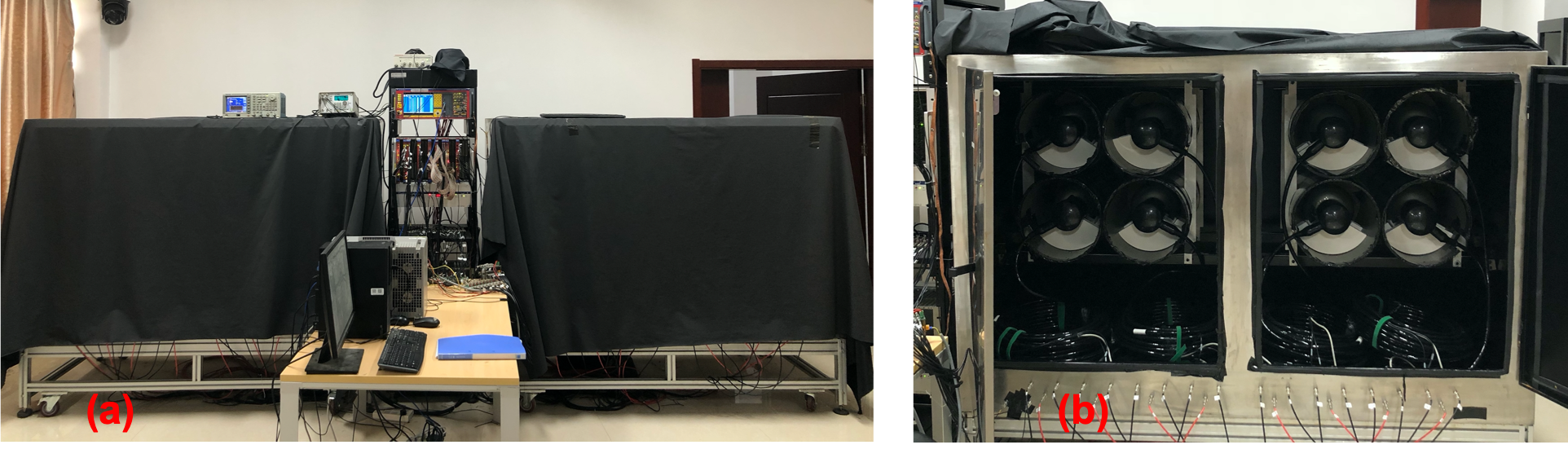}
\vspace*{+0mm}
\caption{The PMT test system.} \label{fig:facility}
\end{center}
\end{figure}
}
\subsection{Setup of the test system}\label{setuptestfacility}
The test system is shown in Fig.~\ref{fig:facility}. Two dark boxes are used with inner walls painted black. Each dark box can hold 8 PMTs (as shown in Fig.~\ref{fig:facility}(b)). The PMTs are put into black cylinders of the racking system with holders specially designed for the CR365-02-1. The cylinder has a diameter of 23 cm and a height of 80 cm. Along the axis of the cylinder and on the back wall of the dark box, there is an optic fibre illuminating the PMT with light from the light sources. The distance between the PMT's photocathode and the fibre is kept to about 40 cm to illuminate the entire photocathode, based on the photon density distribution tested previously. The light is isolated by the black cylinder. The cylinder is wrapped by a $\mu$-metal to shield the Earth's magnetic field (EMF). To minimize the EMF effect, we place the dark boxes north-south so that the $\mu$-metal cylinders are east-west. After shielding, the magnetic field is about 3.0 $\mu$T along the vertical direction, 4.0 $\mu$T along the north-south direction, and 2.0 $\mu$T along the east-west direction as measured by a handy Gauss meter. The EMF along the direction perpendicular to the axis of the $\mu$-metal cylinder is reduced to about one-tenth. 
The 30-meter-long cables are put at the bottom of the dark boxes and are connected to the SHV or BNC connectors assembled at the front wall of the dark boxes. An interlock system is implemented, which automatically turns off the high voltage if any door of the dark boxes is open. 

\subsection{The photon generation system}\label{lightsources}
The light sources consist of two blue LEDs (Hebei 510LB7C, 460-475 nm) and a pico-second laser pulser (Hamamatsu PLP-10, 405 nm).
The laser pulser is used to measure the SPE spectrum and transit time spread (TTS) with the intensity tuned to the SPE level.
The LEDs driven by a pulse generator (Tektronix AFG3252) are used for other measurements such as afterpulse rate and linearity.
To illuminate 16 PMTs at the same time, a light mixing and distributing system has been developed (as shown in Fig.~\ref{fig:lightsource}(a)). 
{
\begin{figure}[htbp]
\begin{center}
\includegraphics[keepaspectratio,width=0.90\textwidth]{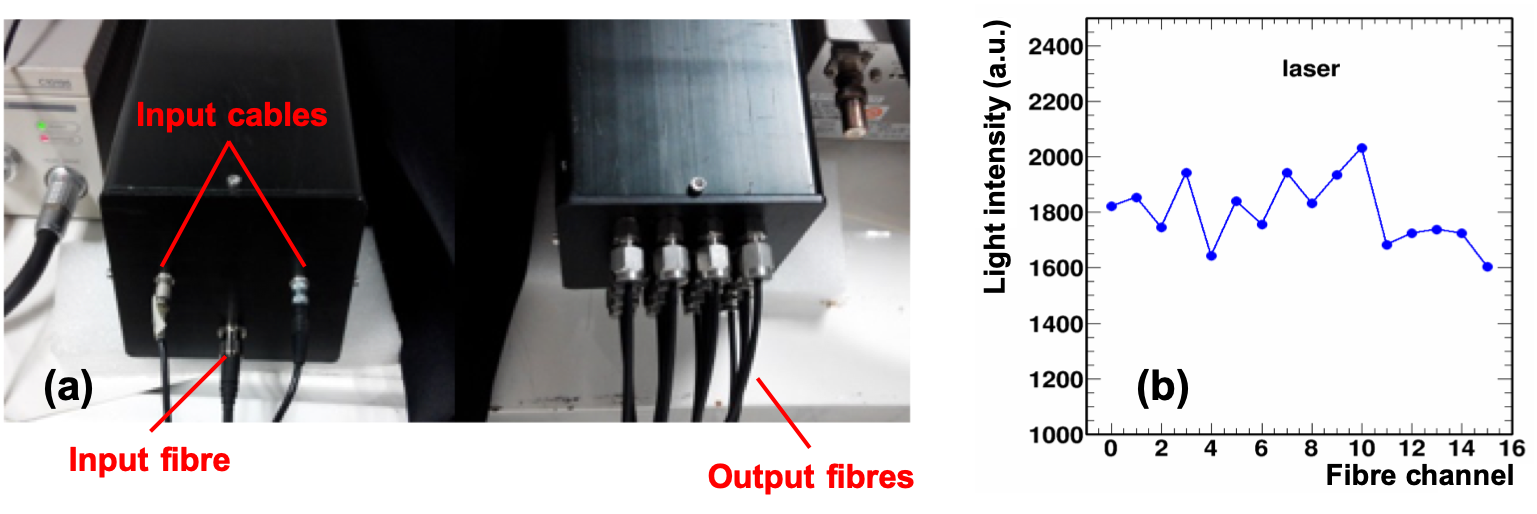}
\vspace*{+0mm}
\caption{(a) The light distribution system. (b) The relative light intensities of the 16 selected fibres of the light distribution system.} \label{fig:lightsource}
\end{center}
\end{figure}
}
The inputs of the light mixing and distributing system include: an optic fibre transmitting the light from the laser pulser and two coaxial cables sending driver signals for two LEDs. Inside the system, the optic fibre and two LEDs are attached to one end of a hexagonal light pipe with a certain incident angle to mix and homogenize the light from the laser pulser and LEDs. On the other end of the light pipe, a bundle of 20 optic fibres are attached to distribute the light.  
16 out of the 20 fibres are chosen to illuminate the 16 PMTs and the other 4 are spares.
Fig.~\ref{fig:lightsource}(b) shows the relative light intensities of the 16 selected fibres. 
The uniformity of the light intensities is within $\pm10\%$. A good uniformity of light intensities is essential to test 16 PMTs at the same time.

\subsection{The Electronics System }\label{DAQ}
Figure~\ref{fig:daq} shows the schematic diagram of the electronics system.  
A BNC575  digital pulse generator from Berkeley Nucleonics Corp. is equipped with four independent channels. 
Channel A is fed into a Fan-In-Fan-Out (FIFO, CAEN N625) to convert TTL pulses to NIM pulses and then is fed into a dual timer (CAEN N93B). 
The outputs of the dual timer provide the trigger signal for a TDC (CAEN V1290A) and the gate signals for three QDCs.
Channel B is used to trigger a pulse generator (Tektronix AFG3252), generating pulses to drive the two LEDs. 
Channel C provides the time reference for the measurement of the afterpulse rate.
Channel D is used to trigger the pico-second laser pulser.
The synchronous output of the laser provides the time reference for the measurement of TTS. 
{
\begin{figure}[htbp]
\begin{center}
\includegraphics[keepaspectratio,width=0.9\textwidth]{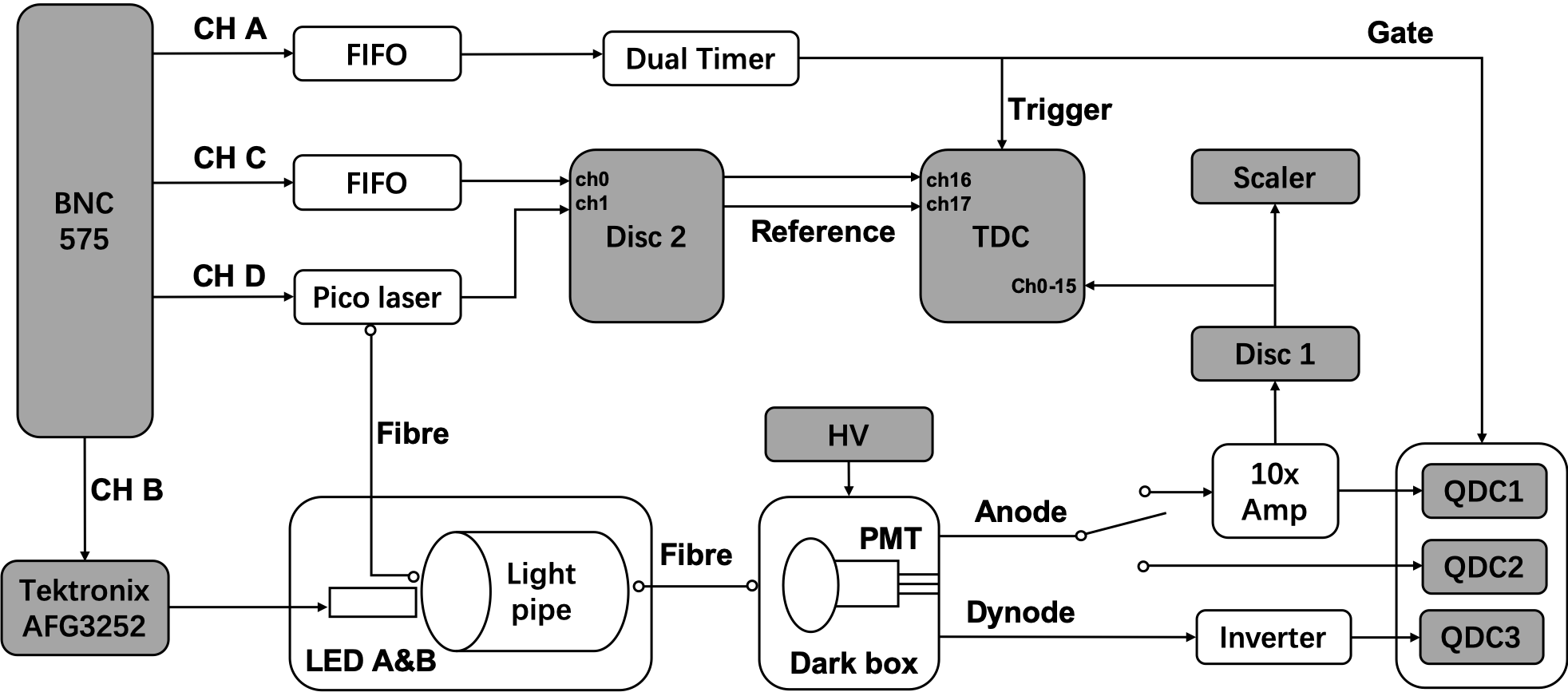}
\vspace*{+0mm}
\caption{Schematic diagram of the data acquisition system. Modules communicated with the DAQ computer are shown in grey color.} \label{fig:daq}
\end{center}
\end{figure}
}
The PMTs are powered by CAEN A1535P in a SY2527 system and are illuminated by the light from the photon generation system.
The anode output of the PMT is connected by a custom made multi-channel switch. 
Each channel of the switch has one input and two outputs: one output is connected to a 10$\times$ fast amplifier (CAEN N979); the other output is connected to QDC2 (CAEN V792N) to measure the linearity of the anode output.
Each channel of the 10$\times$ fast amplifier has two outputs: one output is sent to QDC1 (CAEN V965) to measure the SPE spectrum; the other output is fed into Discriminator1 (CAEN V814).
Each channel of Discriminator1 has two outputs: one output is sent to the TDC (CAEN V1290A) for the measurements of the TTS or afterpulse rate depending on the configuration of the BNC575 and TDC; the other output is fed into a scaler (CAEN V830) for the measurement of the dark count rate. 
The dynode output is fed into a custom made inverter and then into QDC3 (CAEN V965) for the measurement of the linearity of the dynode output.
QDC2 and QDC3 together are used for the measurement of the anode-to-dynode charge ratio.

Fig.~\ref{fig:multiplexer_inverter} shows the schematics of one channel of the switch and inverter.
The switch is implemented through a SPDT toggle switch.
The inverter is based on an operational amplifier AD8000.
{
\begin{figure}[htbp]
\begin{center}
\includegraphics[keepaspectratio,width=0.9\textwidth]{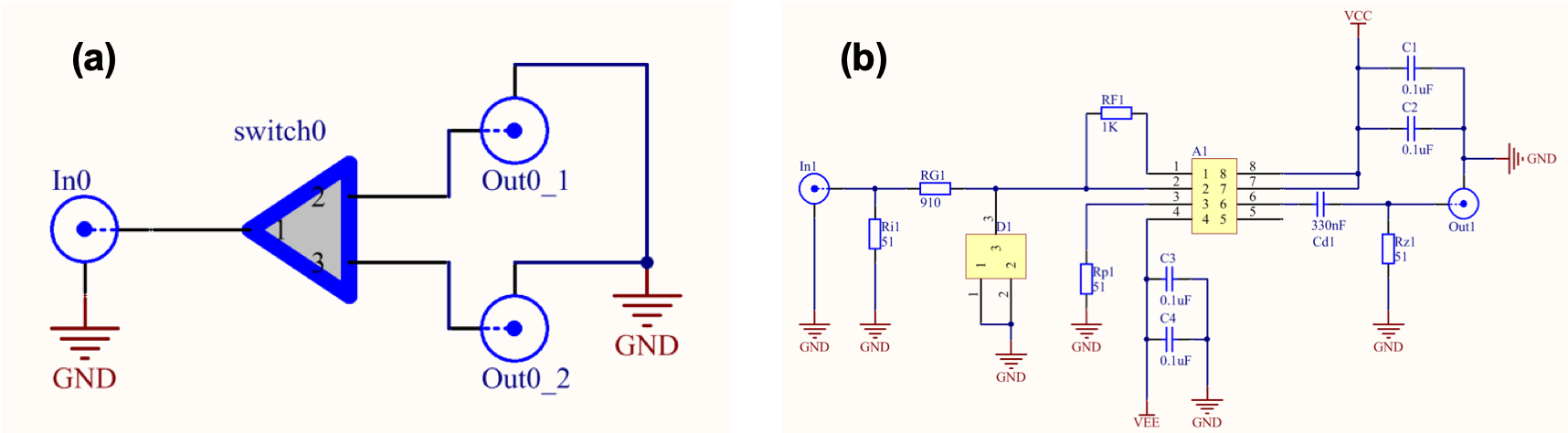}
\vspace*{+0mm}
	\caption{The schematic of the switch (a) and inverter (b) for one channel.} \label{fig:multiplexer_inverter}
\end{center}
\end{figure}
}

\subsection{The Control and Data Acquisition (DAQ) System }\label{Control}
The modules BNC575, pulse generator Tektronix AFG3252, low threshold discriminator, scaler, QDC, TDC, and HV system SY2527 are controlled by a computer via LabVIEW (as shown in grey color in Fig.~\ref{fig:daq}). The BNC575 is controlled to generate different patterns of digital pulses on different channels to trigger different light sources and to provide the trigger and gate signals for the electronics. The frequency, amplitude and width of the outputs of the pulse generator Tektronix AFG3252 can be controlled to drive the LEDs. The high voltage supply from SY2527 system is controlled and monitored. The low threshold discriminator, scaler, QDC and TDC communicate with the DAQ computer through the VME bus. 

\section{The test procedure}\label{Procedure}
At the beginning of each run, all 16 PMTs are assembled and the switches select the 10$\times$ fast amplifiers. 
Typically, assembly of the PMTs takes place in the afternoon making it convenient to allow the PMTs to sit overnight for HV training. Thus, the PMTs sit in the dark boxes for at least 12 hours. During this period, a high voltage of 1200V is applied to all PMTs. The charge outputs with an illumination of about 30 PEs and the dark count rate are monitored every 15 minutes. It takes about 3 to 4 hours for CR365-02-1 to stabilize (gain variation $<5\%$, dark noise rate variation $<10\%$). After 12-hour HV training, the following PMT parameters are measured automatically: high voltage response, SPE spectrum, TTS, quantum efficiency (QE), dark count rate, and afterpulse rate. These measurements take about 2 hours. After that, the switches select the QDC2 for the measurement of linearity and anode-to-dynode charge ratio, which take about 1 hour.

\section{Measurements and results}\label{measurements}

\subsection{High voltage response}\label{hv}
In this measurement, one of the two LEDs is used. We first set the HV to 1200 V to measure the SPE spectrum.
The light intensity of the LED is adjusted so that about 96\% of the time no PE is received by the first dynode of the PMT. 
Assuming the number of PEs received by the first dynode per trigger is Poisson distributed,
the probability of receiving 1 PE is 3.9\%,  and the probability of receiving more than 1 PE is 0.1\%. 
Fig.~\ref{fig:gethv}(a) shows the SPE spectrum at 1200 V. 
The spectrum is described by the Bellamy et al. model~\cite{Bellamy:1994bv}:
\begin{eqnarray}
	S\left(x\right)=e^{-\lambda}\left\{\frac{1-\omega}{\sqrt{2\pi}\sigma_{ped}}e^{-\frac{\left(x-\mu_{ped}\right)^{2}}{2\sigma_{ped}^2}} + \omega\alpha e^{-\alpha\left(x-\mu_{ped}\right)}\left[0.5+0.5\times Erf\left(\frac{x-\mu_{ped}}{\sqrt2\sigma_{ped}}\right)\right]\right\} \nonumber \\
	+ \sum^4_{n=1}\frac{\lambda^ne^{-\lambda}}{n!}\frac{1}{\sigma\sqrt{2n\pi}}e^{-\frac{\left(x-\mu_{ped}-n\mu\right)^{2}}{2n\sigma^2}}\;,
\label{eq:spe}
\end{eqnarray}
Formula~\ref{eq:spe} describes relatively well the experimental shape and it is used to extract values that do not depend on the physical interpretation of the parameters.
We also try the Dossi et al. model~\cite{Dossi:1998zn} to describe the data. 
In the Bellamy et al. model, the SPE response of a PMT is approximated by a Gaussian distribution, the low-amplitude pulses are considered as noise. 
The Dossi et al. model includes the low-amplitude pulses in the SPE response and models the SPE response as a Gaussian plus an exponential. 
Both models provide the comparable quality fit.
The mean value of the SPE response ($\mu$) extracted from the Bellamy et al. model is $(5828\pm45)$ fC. 
The mean value of the Gaussian part of the SPE response from the Dossi et al. model is $(5821\pm46)$ fC. 
However, the low-amplitude events decrease the mean value of the SPE response by 10-20\% relative to the Gaussian peak~\cite{Dossi:1998zn}. 
{
\begin{figure}[htbp]
\begin{center}
\includegraphics[keepaspectratio,width=0.45\textwidth]{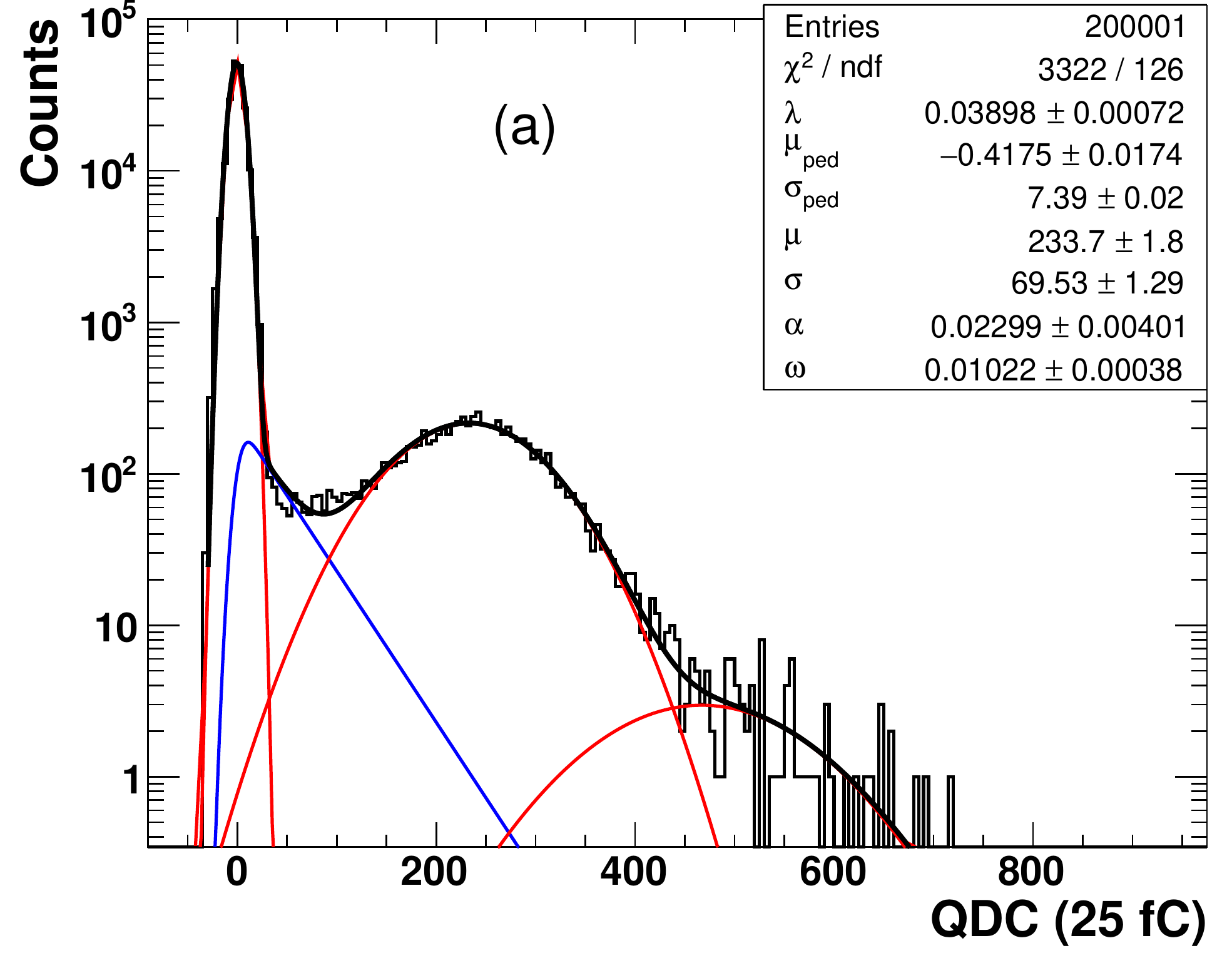}
\includegraphics[keepaspectratio,width=0.45\textwidth]{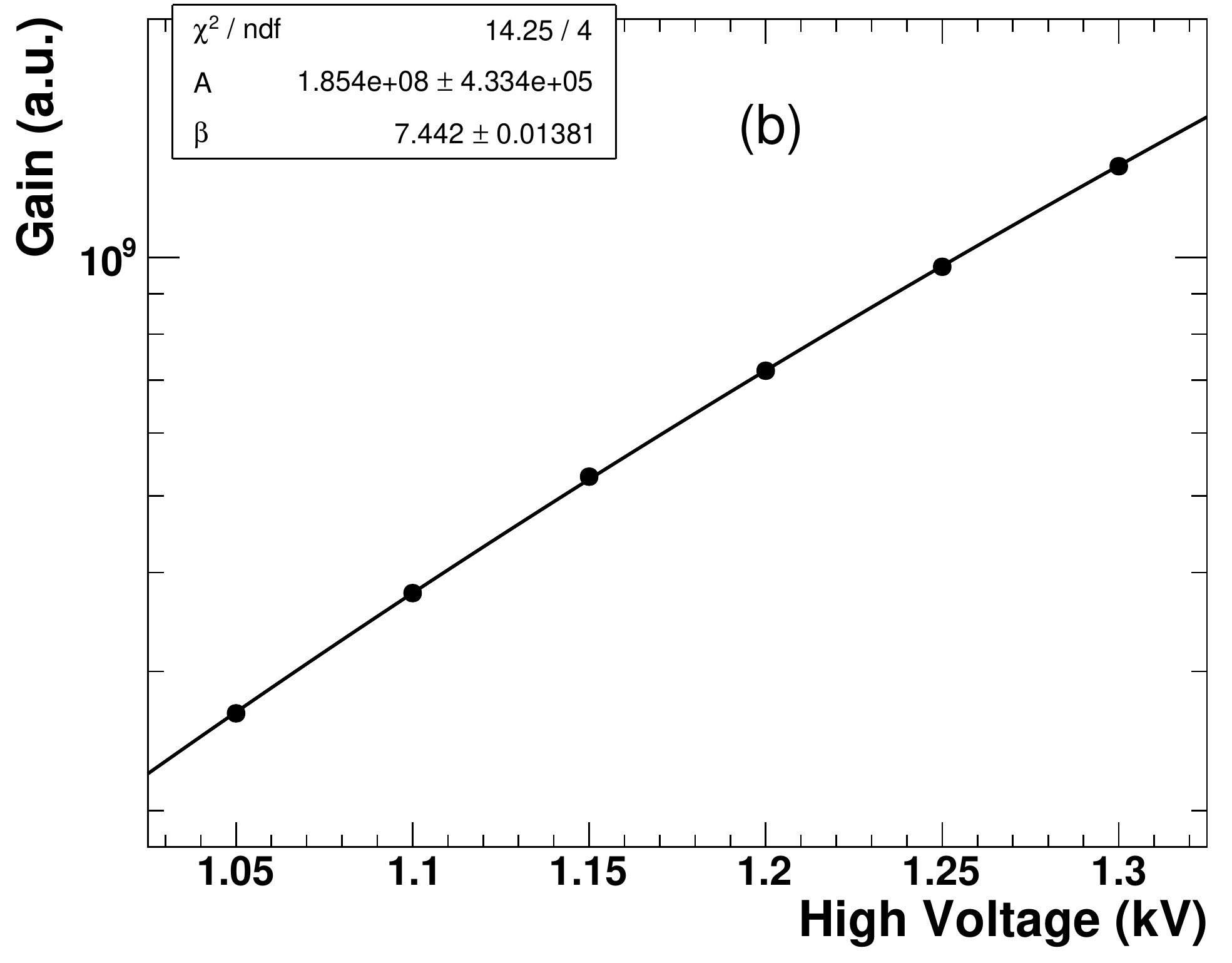}
\vspace*{+0mm}
\caption{(a) The SPE spectrum at 1200V. The red narrow Gaussian distribution on the left is the pedestal. The blue distribution is the convolution of a Gaussian and an exponential function. (b) The charge output as a function of voltage.} \label{fig:gethv}
\end{center}
\end{figure}
}

The measurement of SPE spectrum takes relatively long time because the light intensity needs to be adjusted to the SPE level. 
To speed up, we then increase the light intensity to about 30 PEs to study the HV dependence. Fig.~\ref{fig:gethv}(b) shows the charge output at different voltages from 1050 V to 1350 V with a step of 50 V. 
The voltage dependence of the PMT gain follows a power-law function:
\begin{equation}
G=AV^{\beta},
\label{eq:gain}
\end{equation}
The working voltage ($HV_{0}$) necessary for a gain of $3\times10^6$ can be obtained by
\begin{equation}
	HV_0=HV \left(\frac{3\times10^6}{\mu_{HV}\times LSB_{QDC}\times 1.6\times10^{-19} C\times 10}     \right)^{\frac{1}{\beta}}
\label{eq:hv0}
\end{equation}
where 
$\beta$ is the amplification voltage coefficient,
$HV$ is 1200V, $\mu_{HV}$ is the mean of the SPE response at $HV$ measured by QDC1, $LSB_{QDC}$ is the LSB of QDC1. The rest of the measurements described in this paper are measured under $HV_0$.

Fig.~\ref{fig:hv}(a) shows the distribution of $\beta$. 
The distribution has an average value of 7.29 with a standard deviation of 0.12.
Only one PMT ($\beta=6.2$) is rejected for failing to meet the specification for $\beta$ ($mean\pm0.5$). 
Fig.~\ref{fig:hv}(b) shows the distribution of $HV_0$. 
The mean value of $HV_0$ is 1104 V. 
17 PMTs (1.7\%) failed to meet the specification for $HV_0$ ($mean\pm100$ V).
Due to a relatively long tail on the right-hand side of the $HV_0$ distribution,
all failed PMTs have a $HV_0$ larger than 1204 V.
{
\begin{figure}[htbp]
\begin{center}
\includegraphics[keepaspectratio,width=0.45\textwidth]{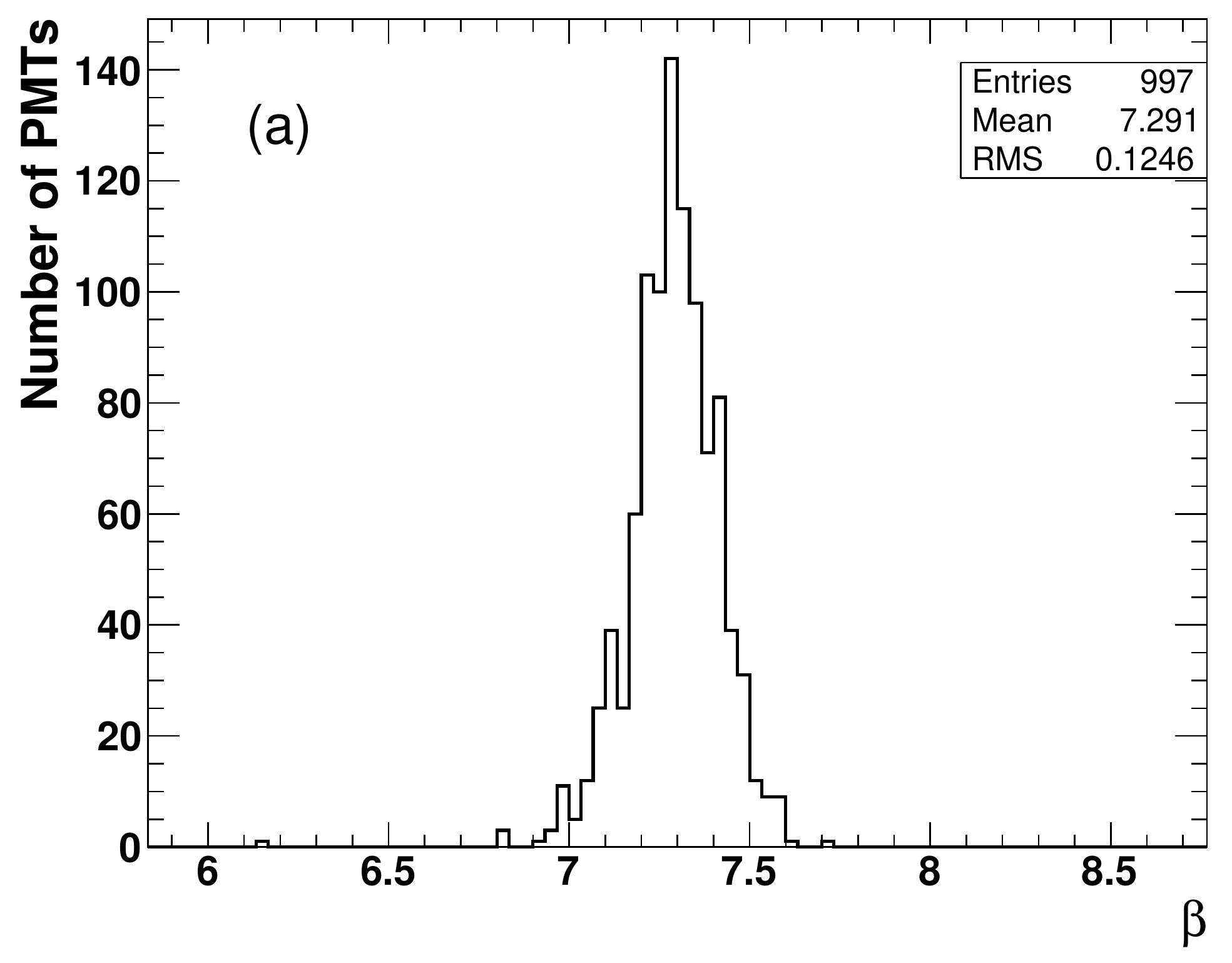}
\includegraphics[keepaspectratio,width=0.45\textwidth]{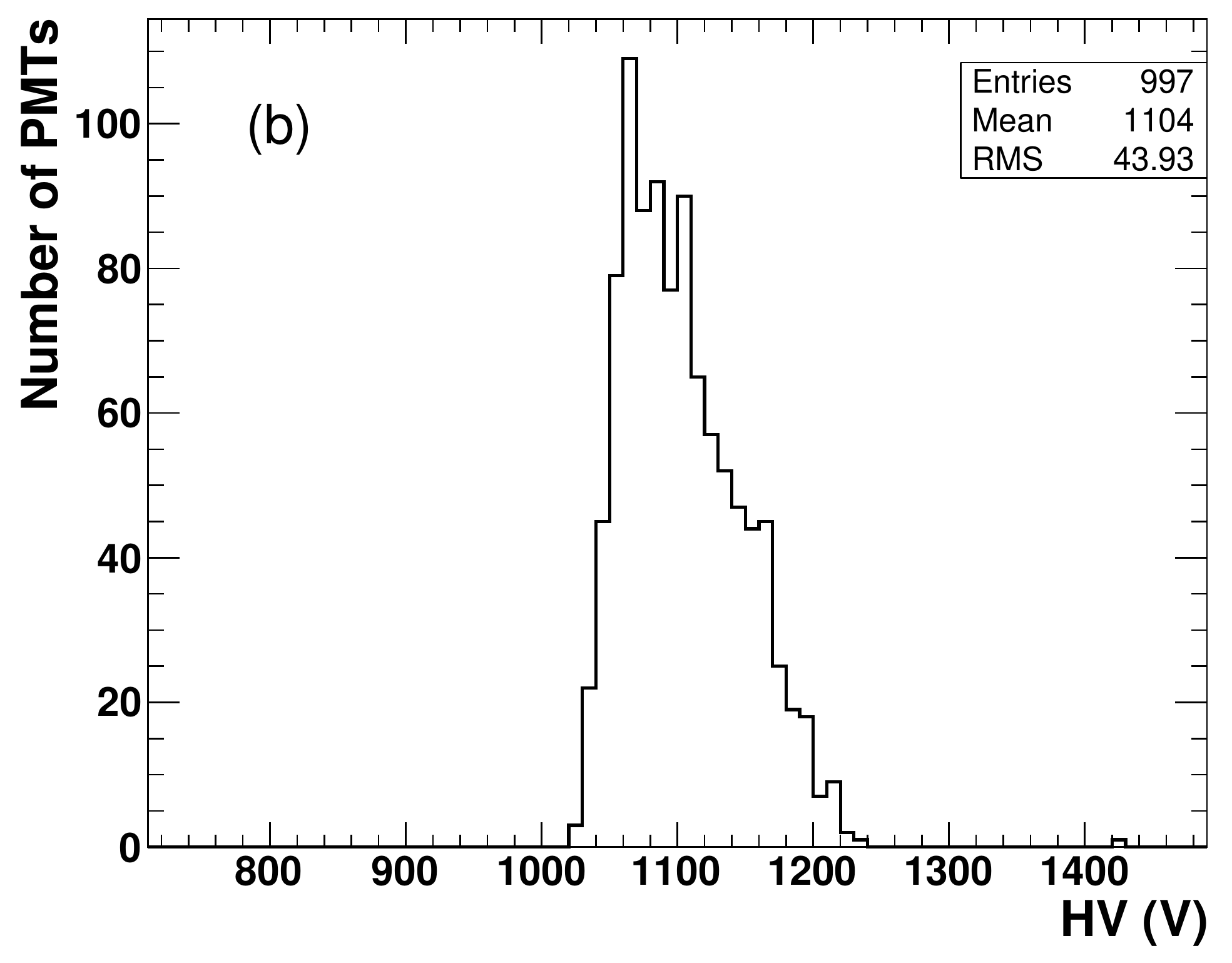}
\vspace*{+0mm}
\caption{(a) Distribution of the amplification voltage coefficient $\beta$. (b) Distribution of the working voltage $HV_0$ for a gain of $3\times10^6$. } \label{fig:hv}
\end{center}
\end{figure}
}

\subsection{SPE response - peak-to-valley ratio, transit time spread and quantum efficiency}\label{spe}
After the measurement of high voltage response, $HV_0$ (Eq.~\ref{eq:hv0}) is obtained. 
Then the supplied high voltage is automatically adjusted to $HV_0$ for each PMT.
The PMTs are illuminated by the pico-second  laser pulser.
The light intensity is adjusted so that the mean number of PEs per trigger is about 0.4.
The charge and time information can be recorded at the same time which enables measuring the charge spectrum of the SPE and transit time spread (TTS) simultaneously.

\begin{figure}[htbp]
\begin{center}
\includegraphics[keepaspectratio,width=0.32\textwidth]{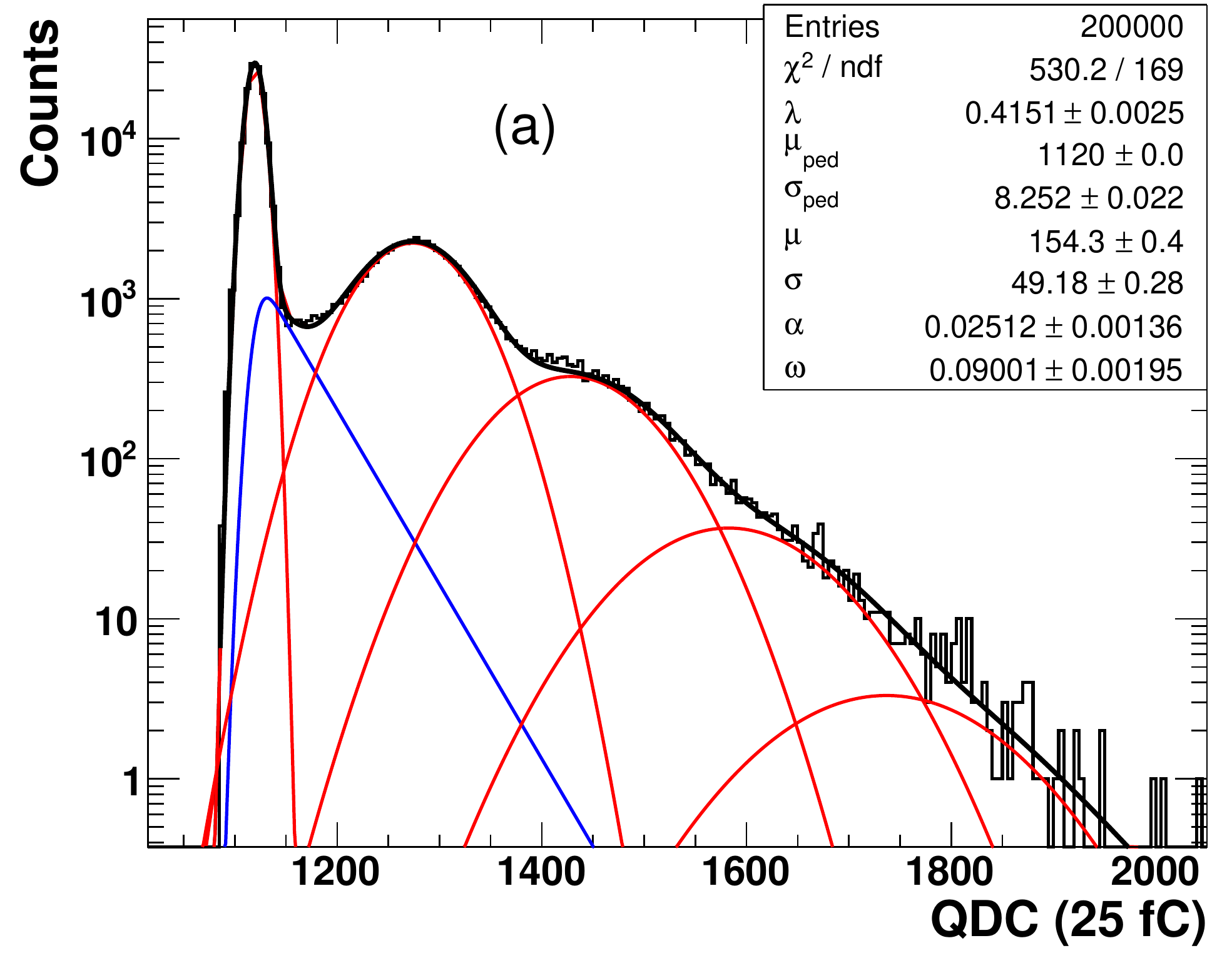}
\includegraphics[keepaspectratio,width=0.32\textwidth]{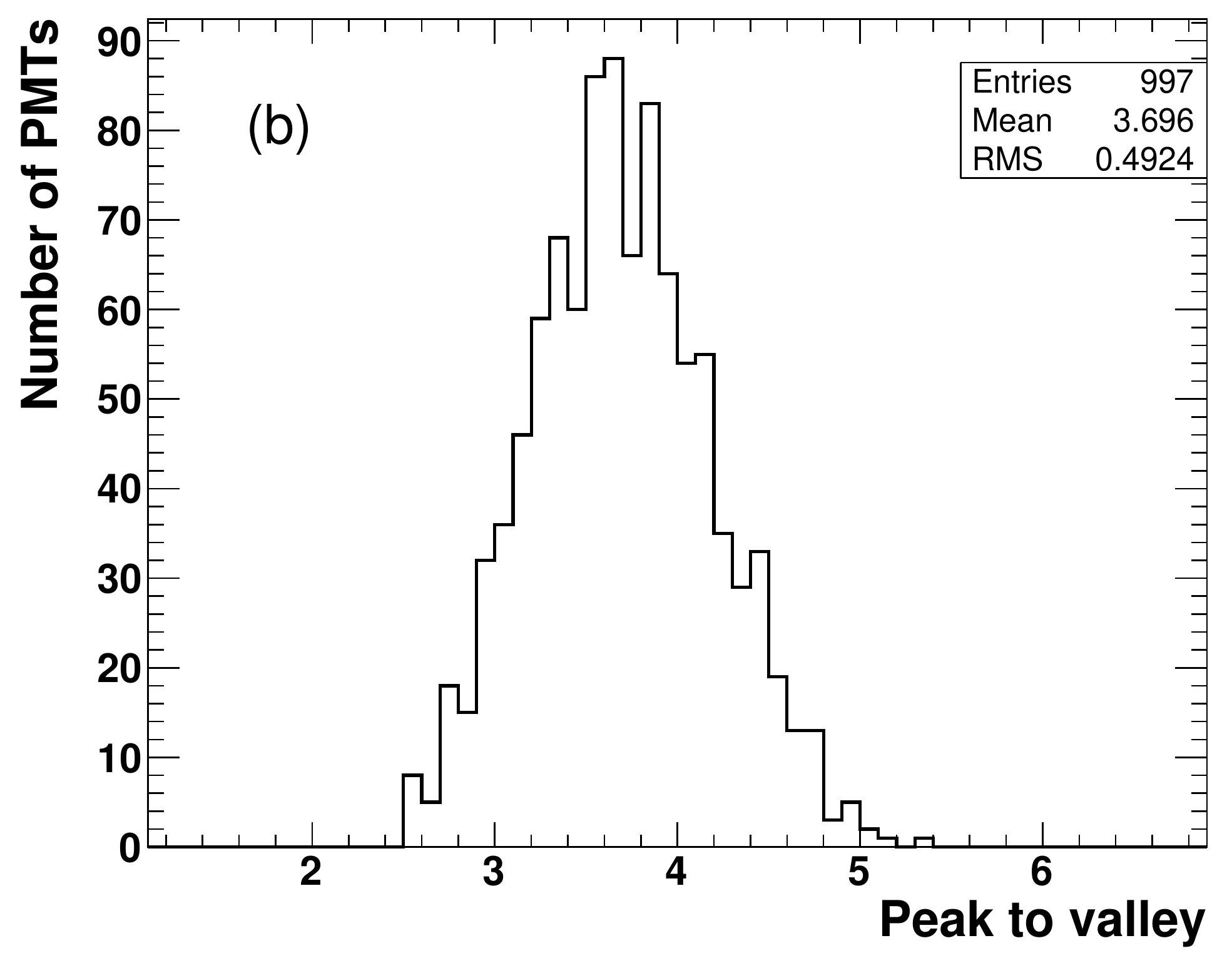}
\includegraphics[keepaspectratio,width=0.32\textwidth]{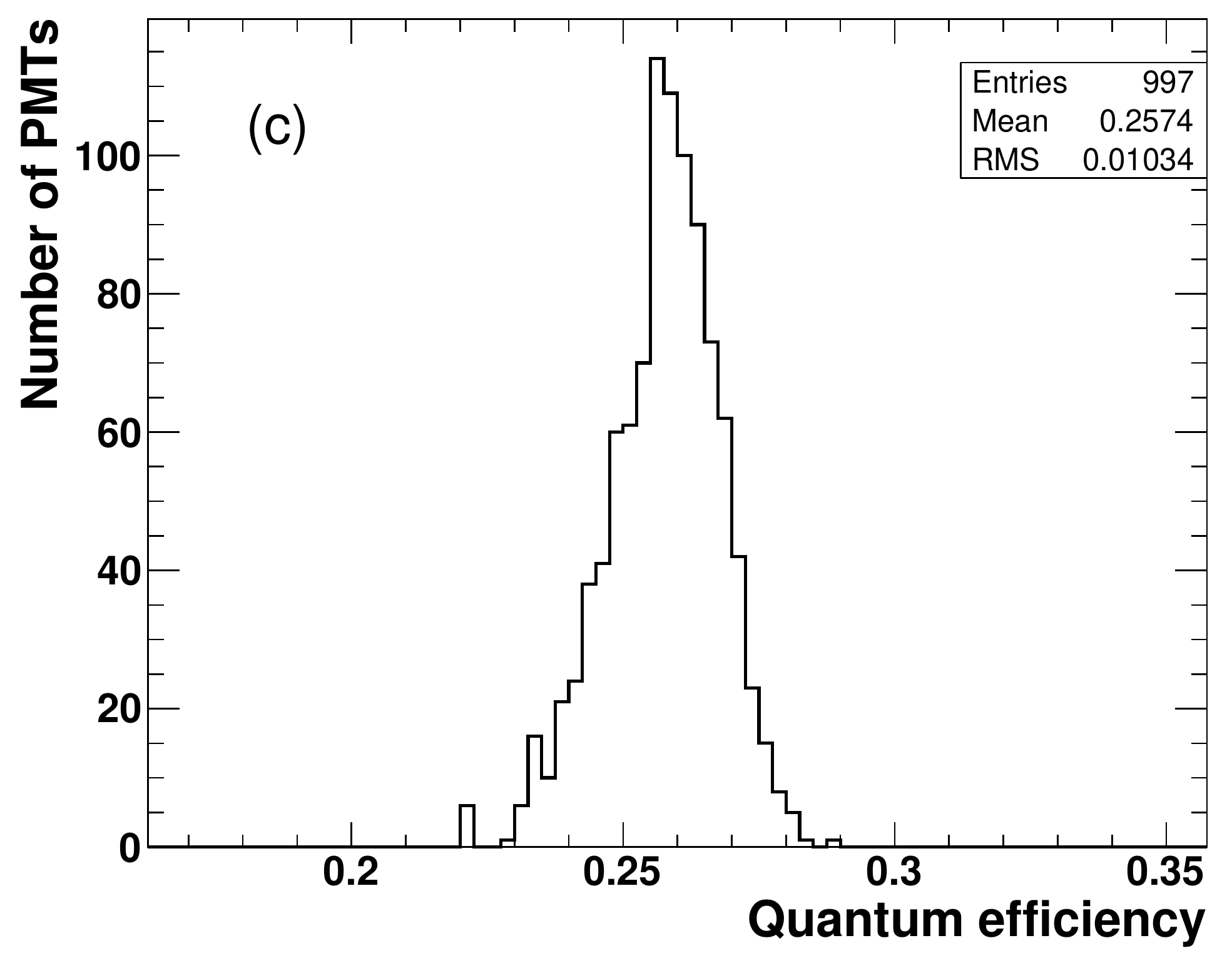}
\vspace*{+0mm}
\caption{(a) SPE spectrum at $HV_0$. (b) Distribution of P/V for all tested PMTs. (c) Distribution of quantum efficiency for all tested PMTs} \label{fig:spehv0_pv}
\end{center}
\end{figure}
Fig.~\ref{fig:spehv0_pv}(a) shows the SPE spectrum at $HV_0$. The spectrum is fitted by Eq.~\ref{eq:spe}.
The peak-to-valley ratio (P/V) is defined as the maximum of the SPE peak divided by the minimum of the valley between the pedestal and the SPE peak.
Fig.~\ref{fig:spehv0_pv}(b) shows the distribution of P/V for all tested PMTs. 
They all meet the requirement of P/V ($>2.0$).

The photon detection efficiency is defined as the product of the quantum- and the collection efficiency. 
The quantum efficiency (QE) is the ratio between the number of produced photoelectrons and the number of incident photons. 
The collection efficiency (CE) is the ratio of the number of electrons collected by the first dynode to the number of produced photoelectrons. 
We use two reference PMTs of which the quantum efficiencies are absolutely calibrated by the BHP, and perform a relative measurement with respect to the reference PMTs. The PMT to be calibrated and the reference PMTs are measured at the same position. 
The QE is obtained by:
\begin{equation}
QE = \frac{SPE}{SPE_{ref}}\times\frac{CE_{ref}}{CE}\times QE_{ref} \;,
\label{eq:rqe}
\end{equation}
where SPE is the number of events in the SPE spectrum with charge larger than $\mu_{ped}+1/3\mu$.
$\mu_{ped}$ and $\mu$ are the mean of the pedestal and the mean of the SPE response, respectively. We have assumed that all PMTs have the same collection efficiency (i.e., $CE_{ref}/CE=1$). According to the information given by the manufacturer, the difference in collection efficiency between PMTs is less than 5\%. A systematic uncertainty of 5\% is therefore introduced. 
Fig.~\ref{fig:spehv0_pv}(c) is the distribution of quantum efficiency. All tested PMTs meet the requirement of QE ($>22\%$).

In the measurement of TTS, the synchronous output of the laser is used as the time reference. The anode signal of the PMT is discriminated by a low threshold discriminator with a threshold of 1/3 PE pulse height. The pulse width of the laser is 85 ps (FWHM).
Fig.~\ref{fig:tts}(a) shows the slewing effect of the arrival time of the SPE signals. A time-charge (T-Q) correction is applied~\cite{Huang:2014yfa}.
{
\begin{figure}[hbtp]
\begin{center}
\includegraphics[keepaspectratio,width=0.32\textwidth]{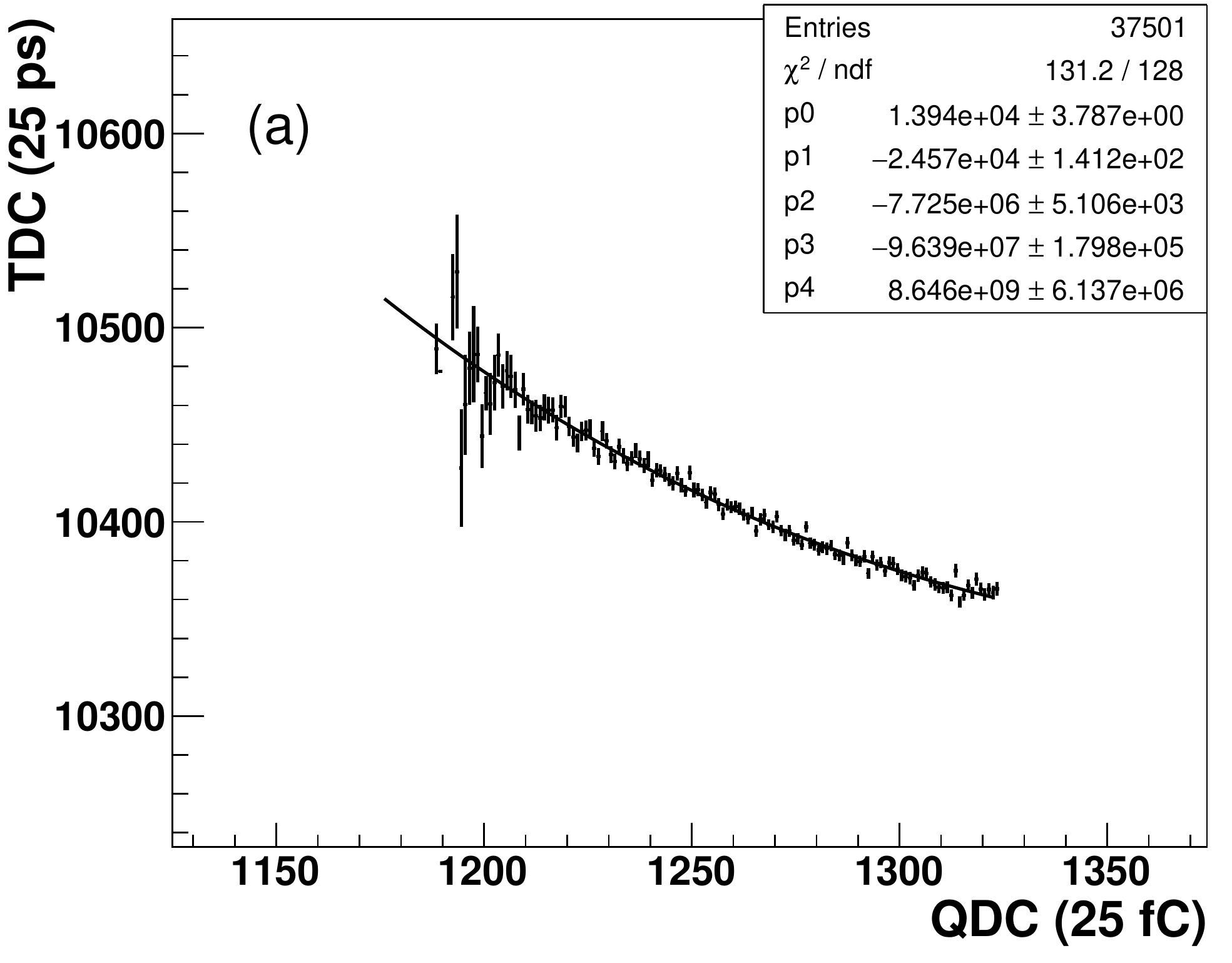}
\includegraphics[keepaspectratio,width=0.32\textwidth]{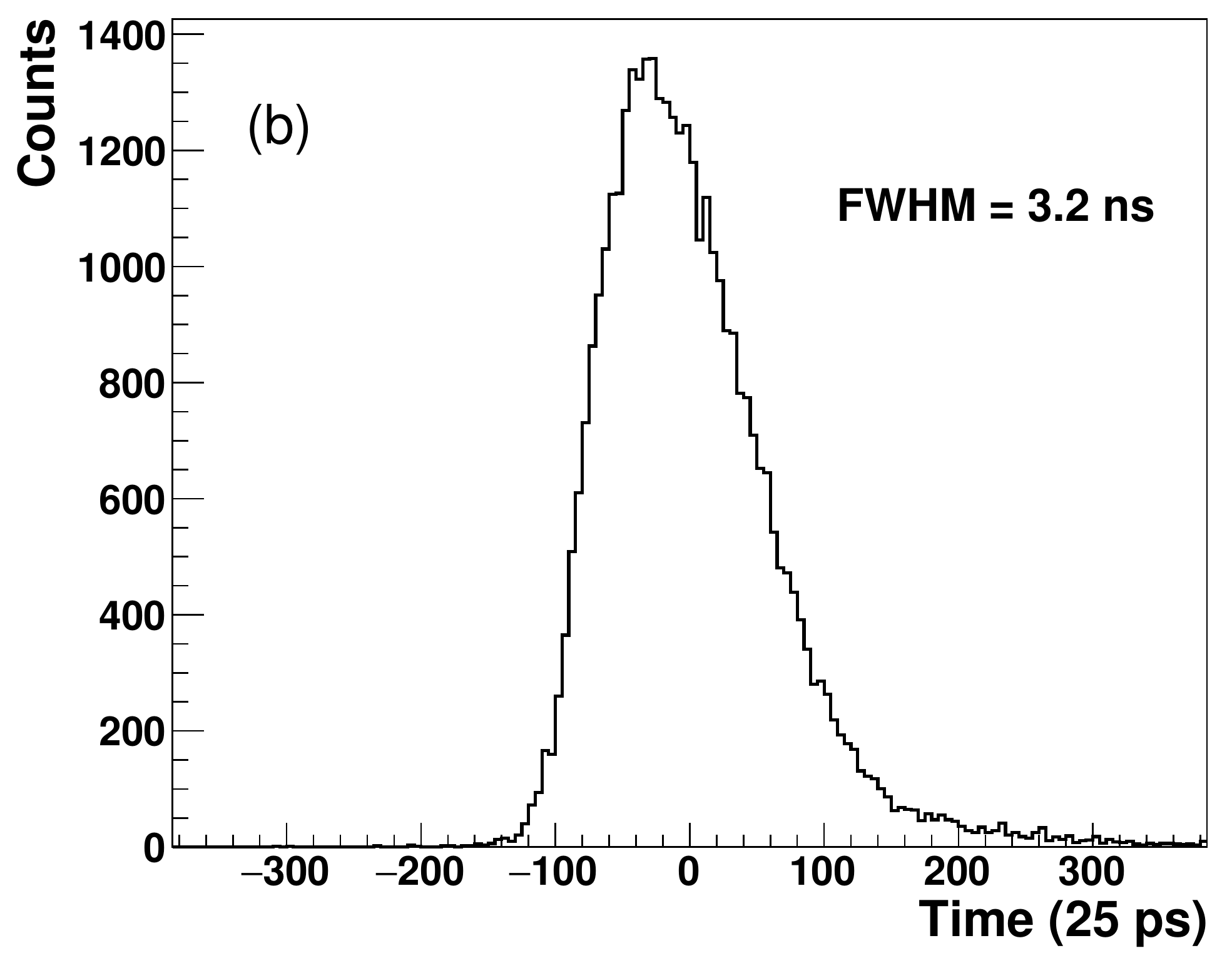}
\includegraphics[keepaspectratio,width=0.32\textwidth]{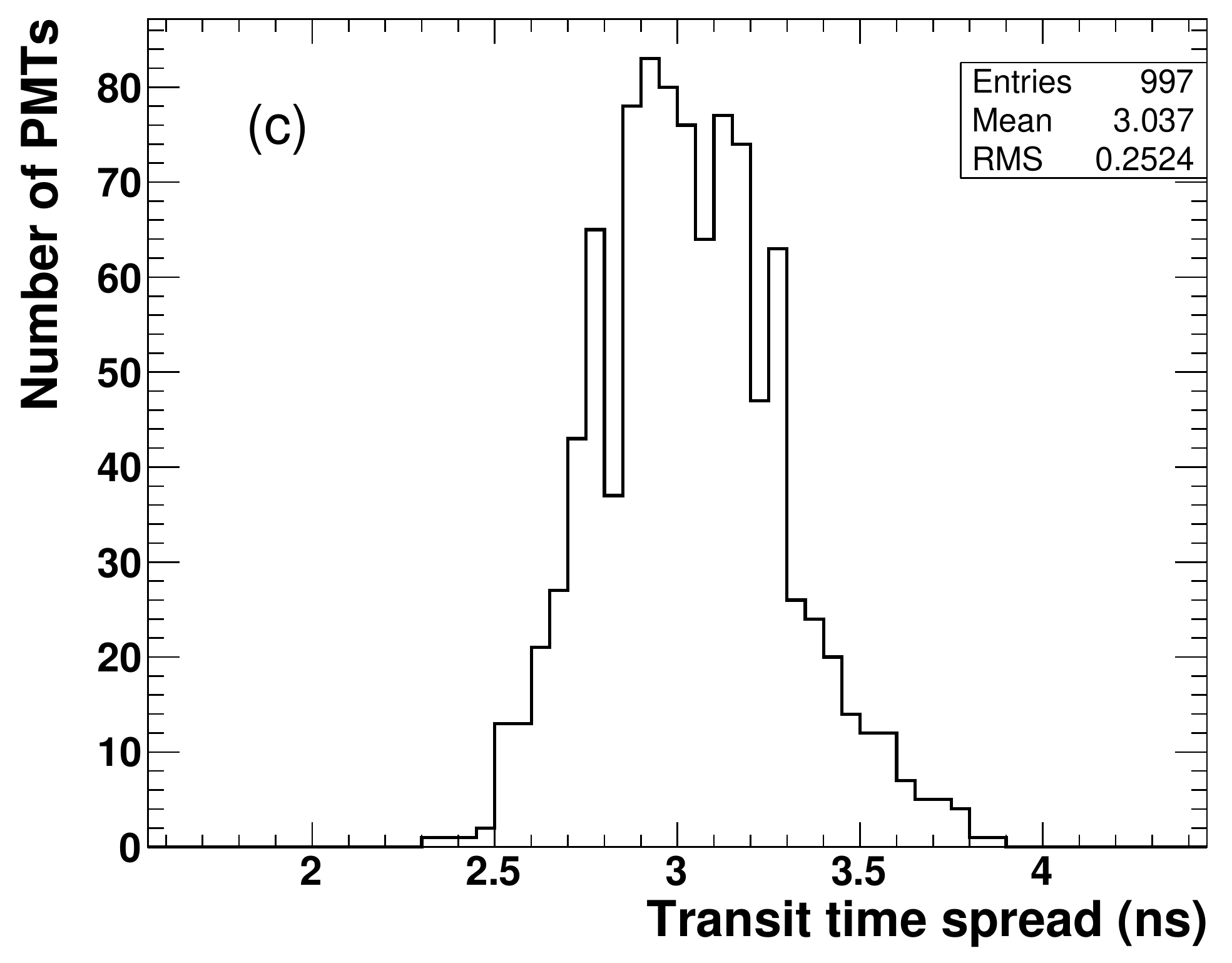}
\vspace*{+0mm}
\caption{(a) The slewing effect of the arrival time of the SPE signals. (b) Distribution of relative transit time after slewing correction. (c) Distribution of TTS for all tested PMTs} \label{fig:tts}
\end{center}
\end{figure}
}
Fig.~\ref{fig:tts}(b) shows the distribution of relative transit time after the slewing correction. 
The TTS is defined as the FWHM of the distribution. 
Fig.~\ref{fig:tts}(c) shows the distribution of TTS for all tested PMTs. 
All tested PMTs meet the requirement of TTS  ($<4$ ns).

\subsection{Dark count rate}\label{dn}
Dark noise will cause accidental coincidences between PMTs.
A high dark count rate will increase the probability of a random trigger in the detector 
and also shift the energy scale of the detector by the amount of the random charge collected during the event window~\cite{Ianni:2004wj}. 
Dark noise is caused mainly by thermal electron emission, leakage current, field emission, scintillation from the glass envelope or electrode support materials, ionization current of residual gases, and noise current caused by cosmic rays and ambient radiations~\cite{pmthandbookv4e}.
Before the measurement of dark count rate, the PMTs are put in the dark boxes for a period of 12 hours. 
The temperature inside the laboratory is stabilized at $20^{\circ}\mathrm{C}$ to $25^{\circ}\mathrm{C}$.
The dark count rate is measured with a threshold of 1/3 PE. 
Fig.~\ref{fig:dn} shows the distribution of dark count rate for all tested PMTs. 
All tested PMTs meet the specification for dark count rate ($<5000$ cps).
{
\begin{figure}[htbp]
\begin{center}
\includegraphics[keepaspectratio,width=0.45\textwidth]{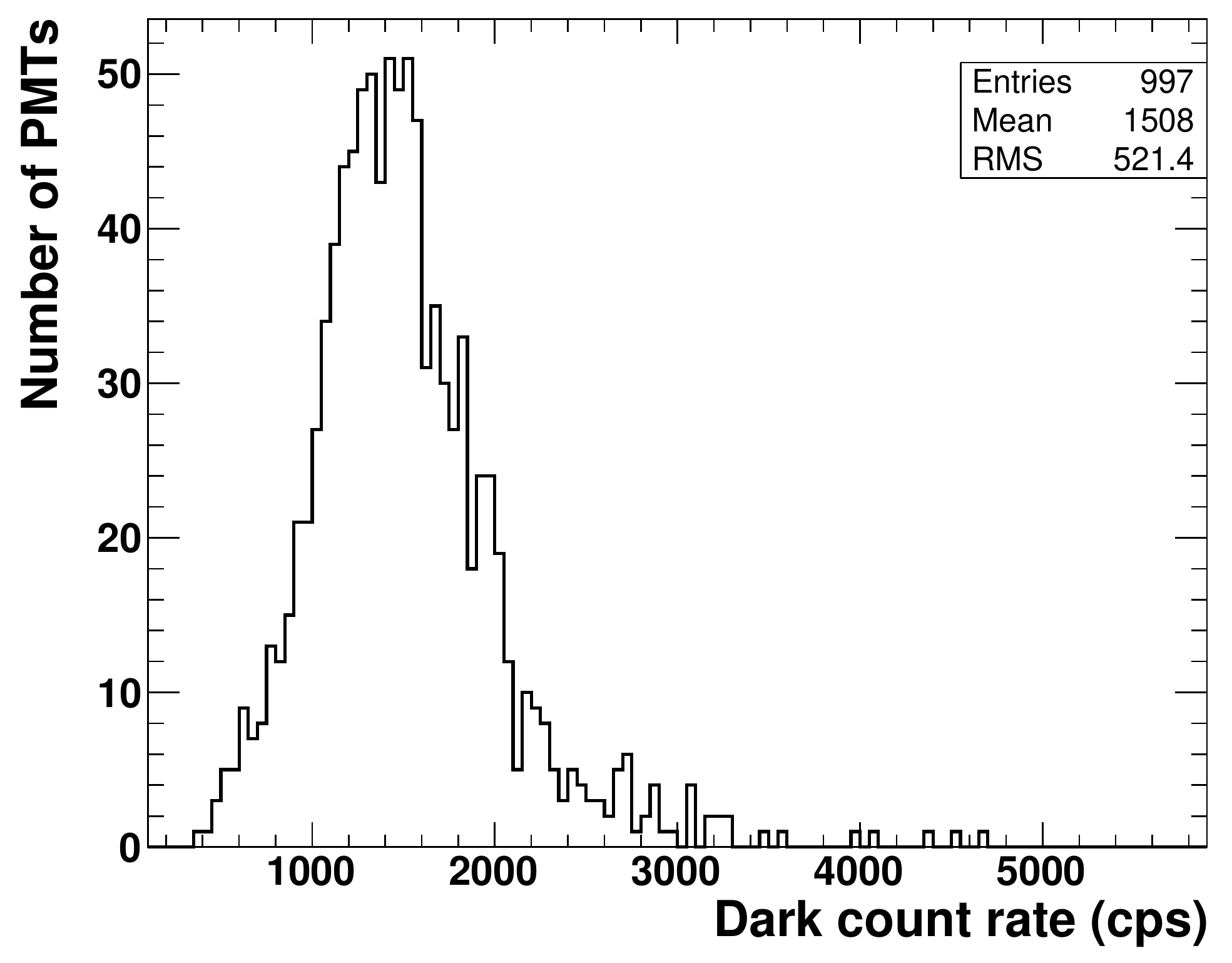}
\vspace*{+0mm}
\caption{The distribution of dark count rate for all tested PMTs} \label{fig:dn}
\end{center}
\end{figure}
}

\subsection{Afterpulse rate}\label{aprate}
Afterpulses are mainly caused by positive ions, which are generated by the ionization of residual gases after being impinged by electrons.
The positive ions will return to the cathode and produce additional electrons.
The time delay of afterpulse is several hundred nanoseconds to a few microseconds with respect to the signal output.
A detailed study of the afterpulse measurement using a multi-hit TDC was introduced in~\cite{Zhao:2016zht}.

In this measurement, PMTs are illuminated by a LED with a light intensity of about 30 PEs. 
The time reference of the TDC is provided by one channel of the BNC generator, while the LED is triggered by another channel.
The time jitter between different channels of the BNC is measured as less than 2 nanoseconds, which is negligible for the measurement of afterpulses over 10 microseconds. 
The PMT signals are discriminated with a threshold of 1/3 PE.
The afterpulse rate is defined as: 
\begin{equation}
	Rate=\frac{N_{Afterpulse}}{A_{Mainpulse}\times N} \times100 \%   \;,
\label{eq:aprate}
\end{equation}
where 
$N_{Afterpulse}$ is the number of afterpulses after subtracting the dark noise counts,
$A_{Mainpulse}$ is the average number of photoelectrons in the main pulse, 
$N$ is the number of events.
Fig.~\ref{fig:aprate}(a) shows a typical afterpulse distribution. 
The main pulse sits at around zero. 
The afterpulse rate between 100 ns and 10 $\mu$s is 0.74\%.
{
\begin{figure}[htbp]
\begin{center}
\includegraphics[keepaspectratio,width=0.32\textwidth]{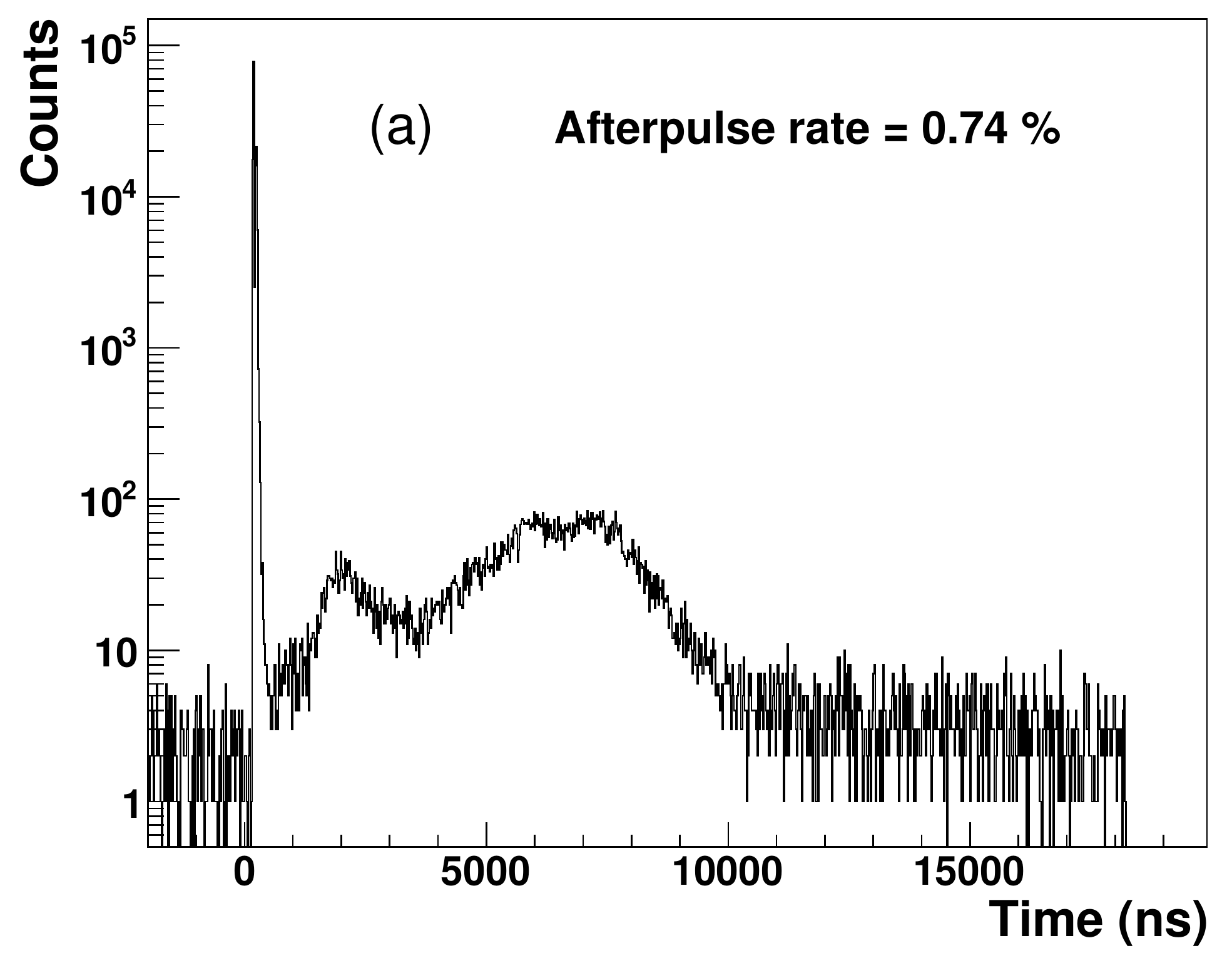}
\includegraphics[keepaspectratio,width=0.32\textwidth]{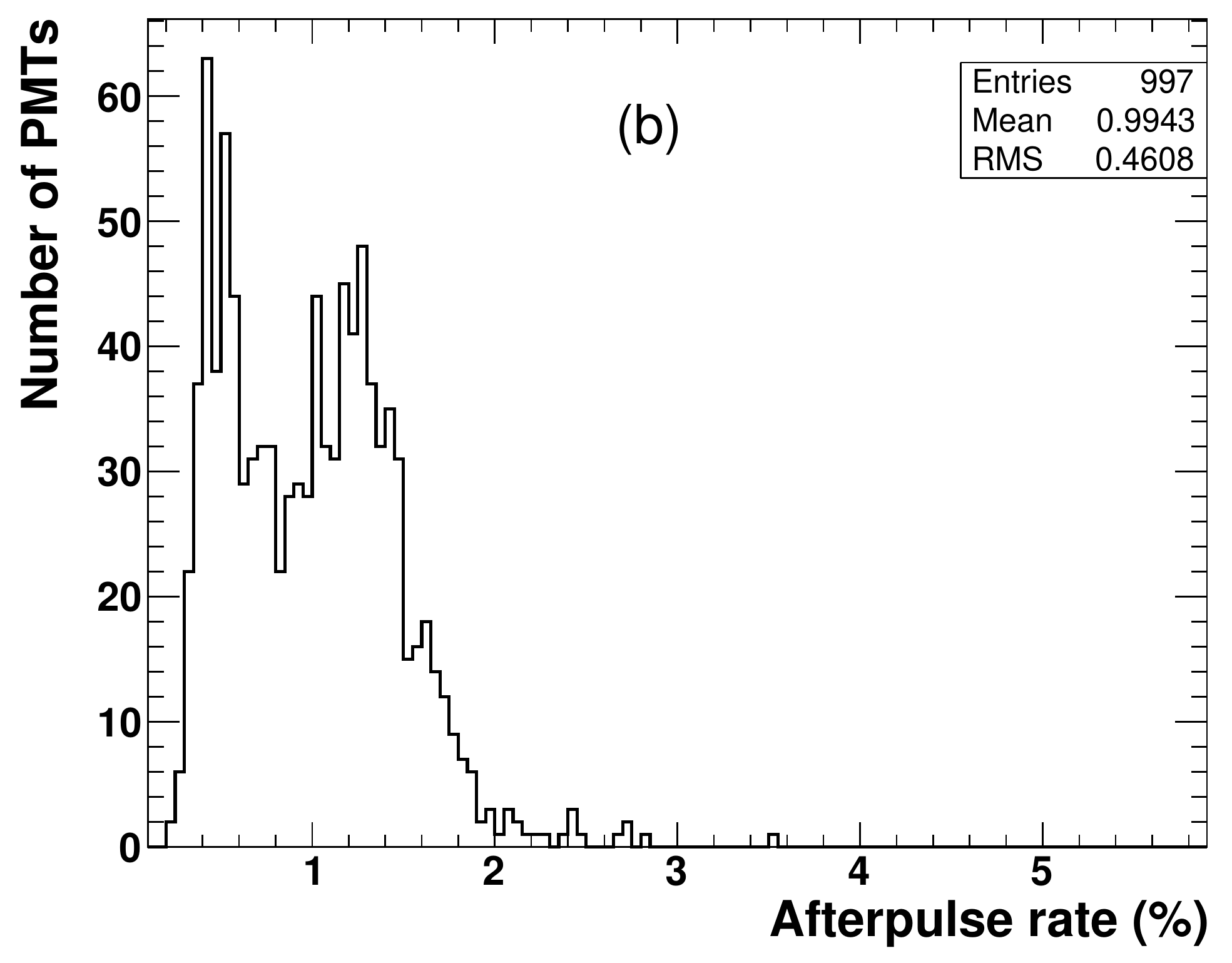}
\includegraphics[keepaspectratio,width=0.32\textwidth]{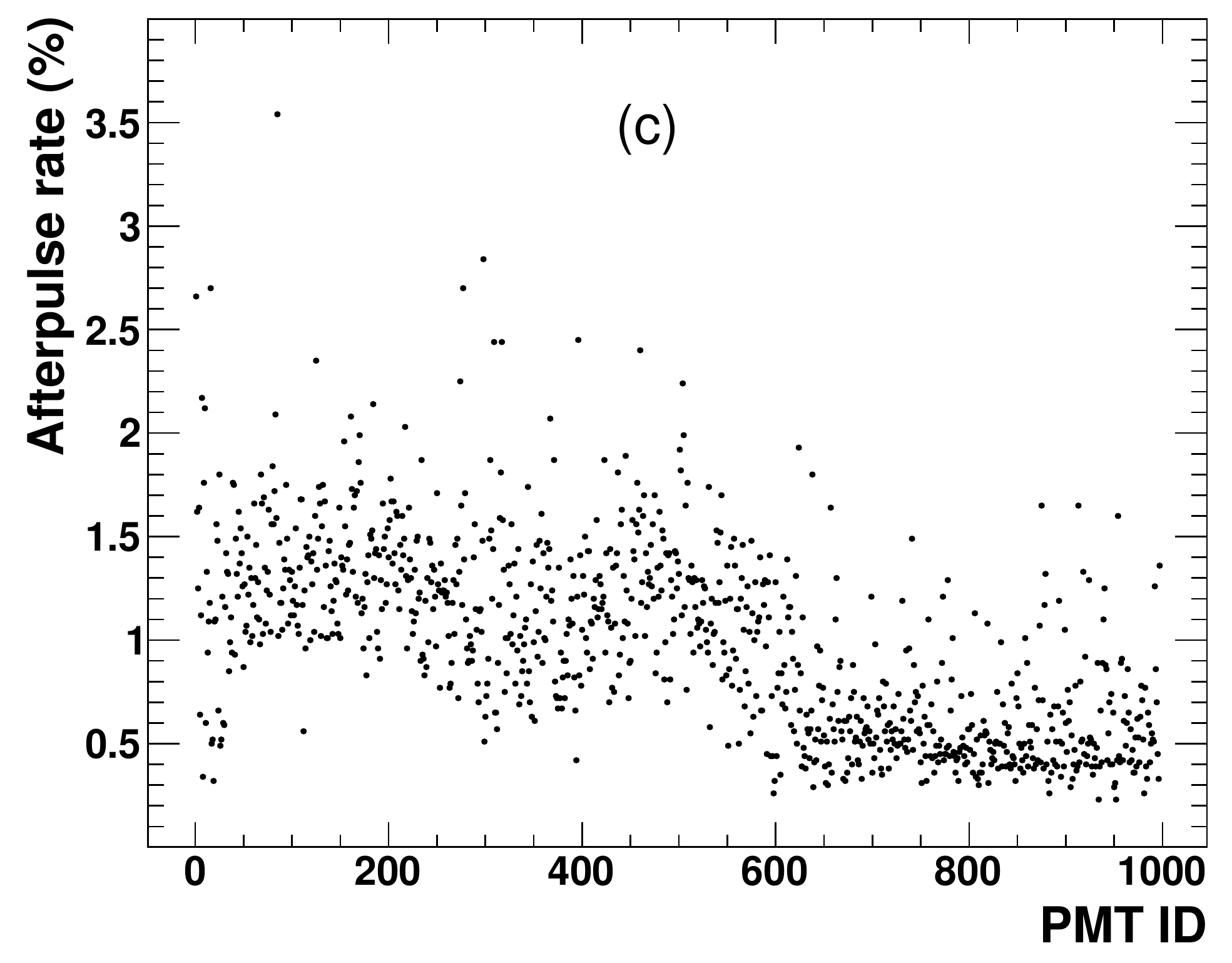}
\vspace*{+0mm}
\caption{(a) A typical afterpulse distribution. The main peak sits at around zero. (b) The distribution of afterpulse rate for all tested PMTs.(c) The afterpulse rate as a function of PMT ID.} \label{fig:aprate}
\end{center}
\end{figure}
}
Fig.~\ref{fig:aprate}(b) shows the distribution of afterpulse rate for all tested PMTs. All PMTs meet the requirement of afterpulse rate ($<5\%$). The distribution of afterpulse rate exhibits a double-peak structure. Fig.~\ref{fig:aprate}(c) shows the afterpulse rate as a function of PMT ID. Roughly the first 600 PMTs produced by the manufacturer have a higher afterpulse rate with a mean value of 1.24\%, which correspond to the second peak in Fig.~\ref{fig:aprate}(b); while the last 397 PMTs have a lower afterpulse rate with a mean value of 0.61\%, which correspond to the first peak in Fig.~\ref{fig:aprate}(b).

\subsection{Nonlinearity and anode-to-dynode charge ratio}\label{nl}
According to the simulation results~\cite{wcdaicrc2011}, a high dynamic range is required for the PMTs used in the LHAASO-WCDA.
To meet the requirements, PMTs are designed to have two outputs: one from the anode, the other from the Dy8.  
A measurement of the charge non-linearity of the anode and the dynode is performed. 
The output charge ratio between the anode and the Dy8 (A/D) has been measured at the same time.

Fig.~\ref{fig:adratio}(a) shows the charge correlation between the anode and the Dy8. The A/D, which is the slope of the curve, is $49.35\pm0.01$. Fig.~\ref{fig:adratio}(b) shows the distribution of A/D for all tested PMTs. The distribution of A/D has an average value of 48.0 with a standard deviation of 3.5. 32 PMTs (3.2\%) failed to meet the specification for A/D ($mean\pm15\%$). The A/D has proven to be the most critical cut in the PMT selection process.
{
\begin{figure}[htbp]
\begin{center}
\includegraphics[keepaspectratio,width=0.45\textwidth]{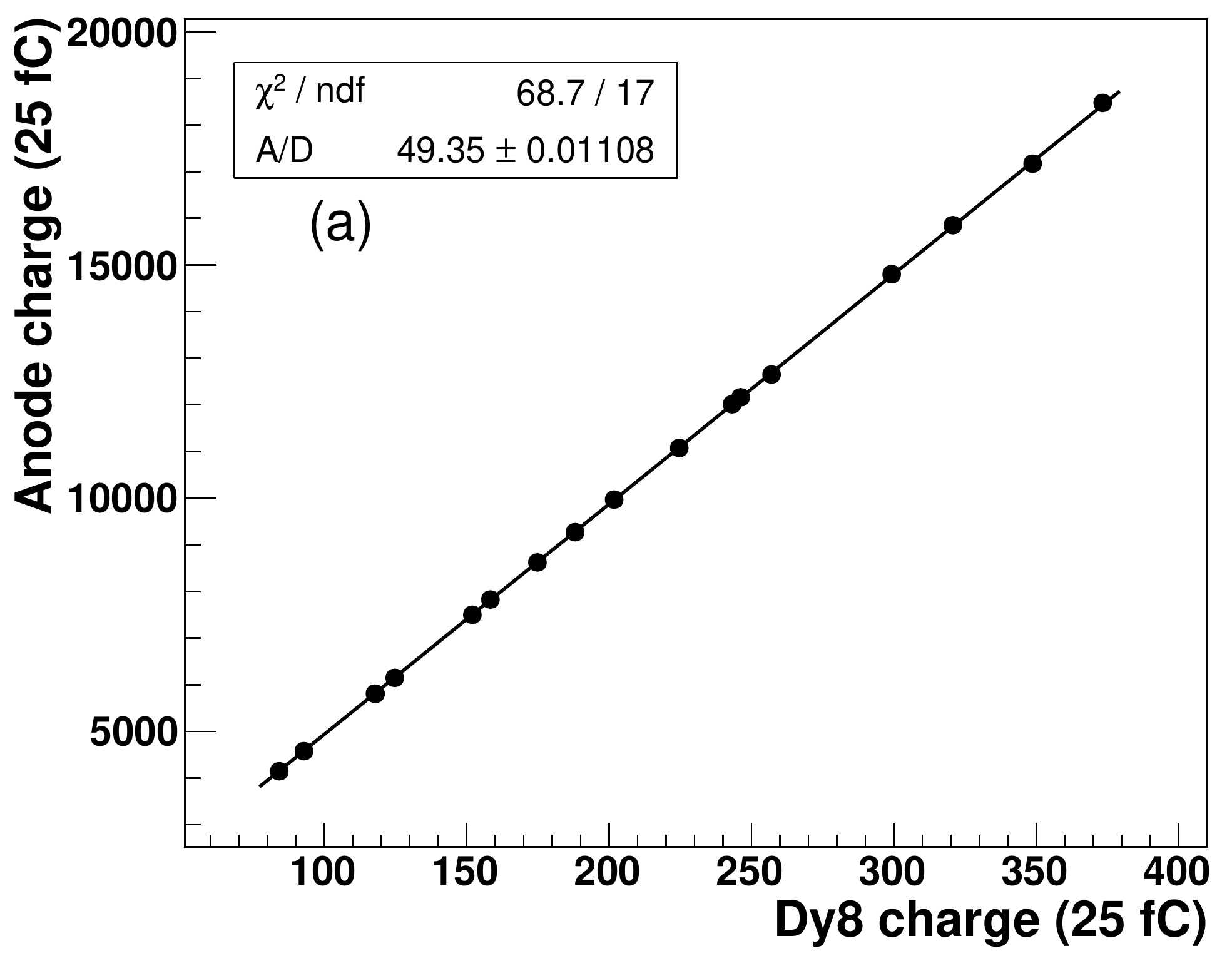}
\includegraphics[keepaspectratio,width=0.45\textwidth]{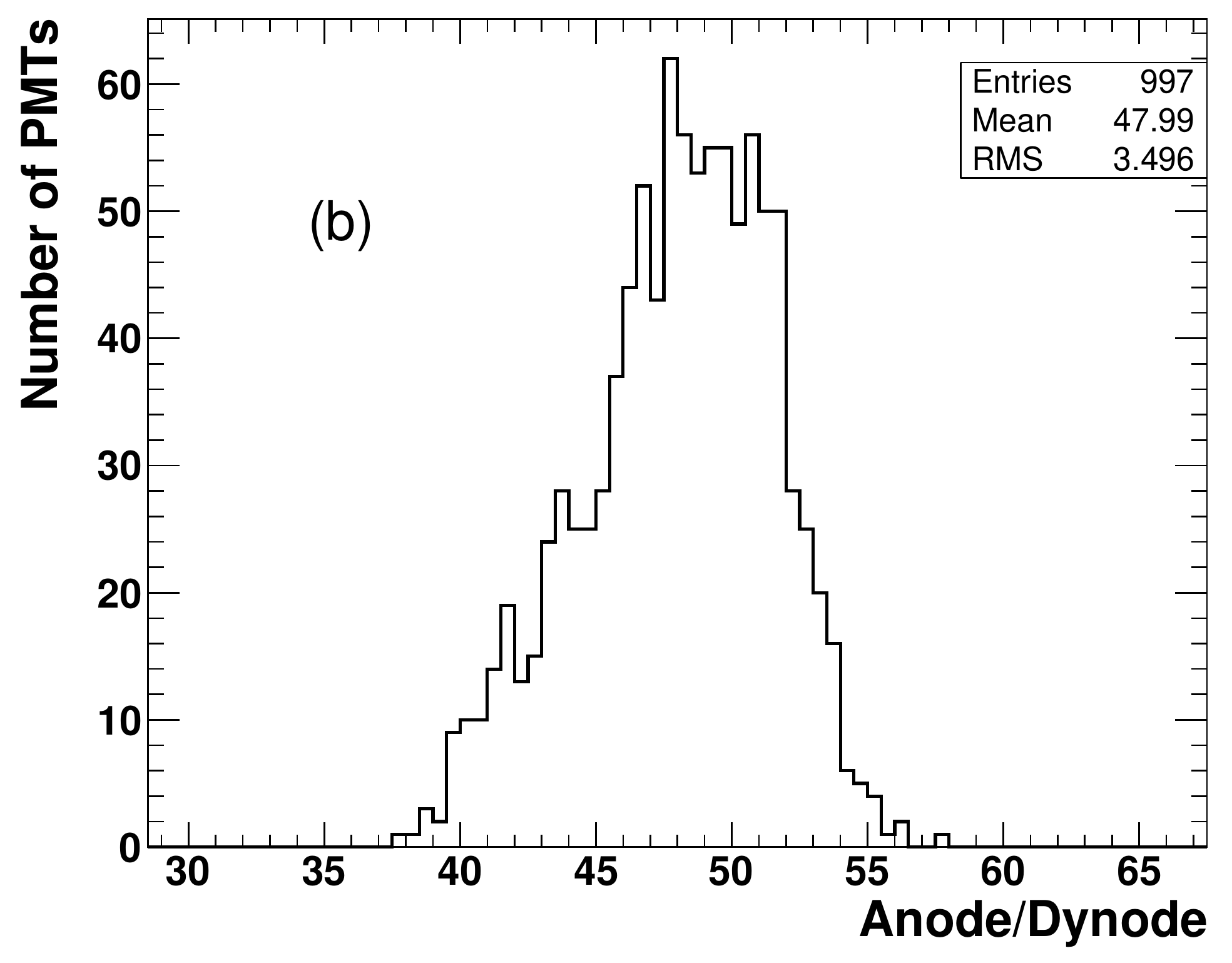}
\vspace*{+0mm}
\caption{(a) The charge correlation between the anode and the Dy8. (b) The distribution of A/D for all tested PMTs.} \label{fig:adratio}
\end{center}
\end{figure}
}

To determine the non-linearity, the A-B method is used~\cite{Barnhill:2008zz,Huang:2013lca}.
The non-linearity is defined as
\begin{equation}
	NL=\frac{Q_{AB}-(Q_A+Q_B)}{Q_A+Q_B} \times100 \%   \;,
\label{eq:nl}
\end{equation}
where $Q_A$ is the charge output from firing LED A alone, 
$Q_B$ is the charge output from firing LED B alone, 
and $Q_{AB}$ is the output from firing LEDs A and B simultaneously.
Fig.~\ref{fig:nl}(a) shows the typical non-linearity curves for the anode and the Dy8 as a function of the number of PEs collected by the first dynode. 
{
\begin{figure}[htbp]
\begin{center}
\includegraphics[keepaspectratio,width=0.45\textwidth]{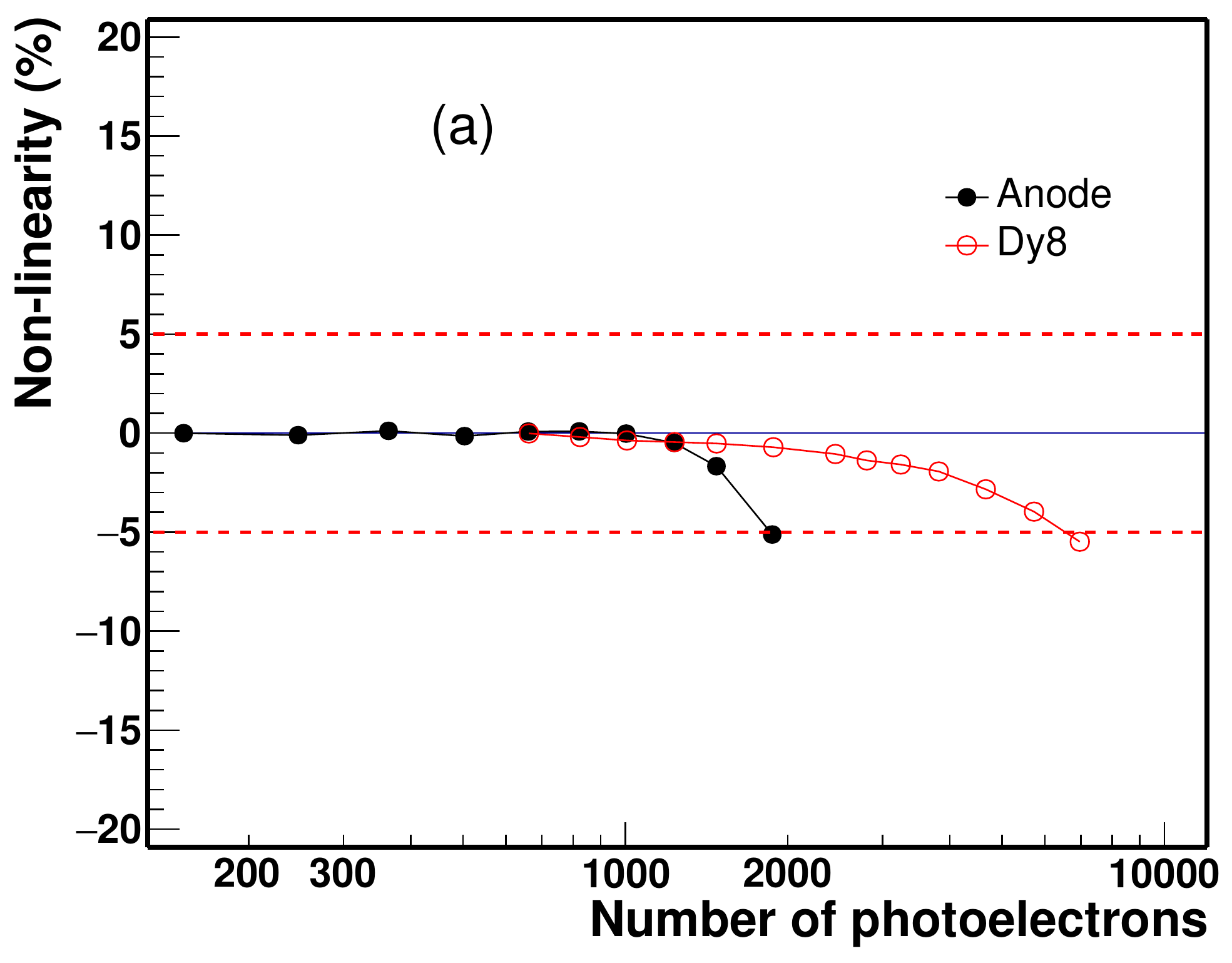}
\includegraphics[keepaspectratio,width=0.45\textwidth]{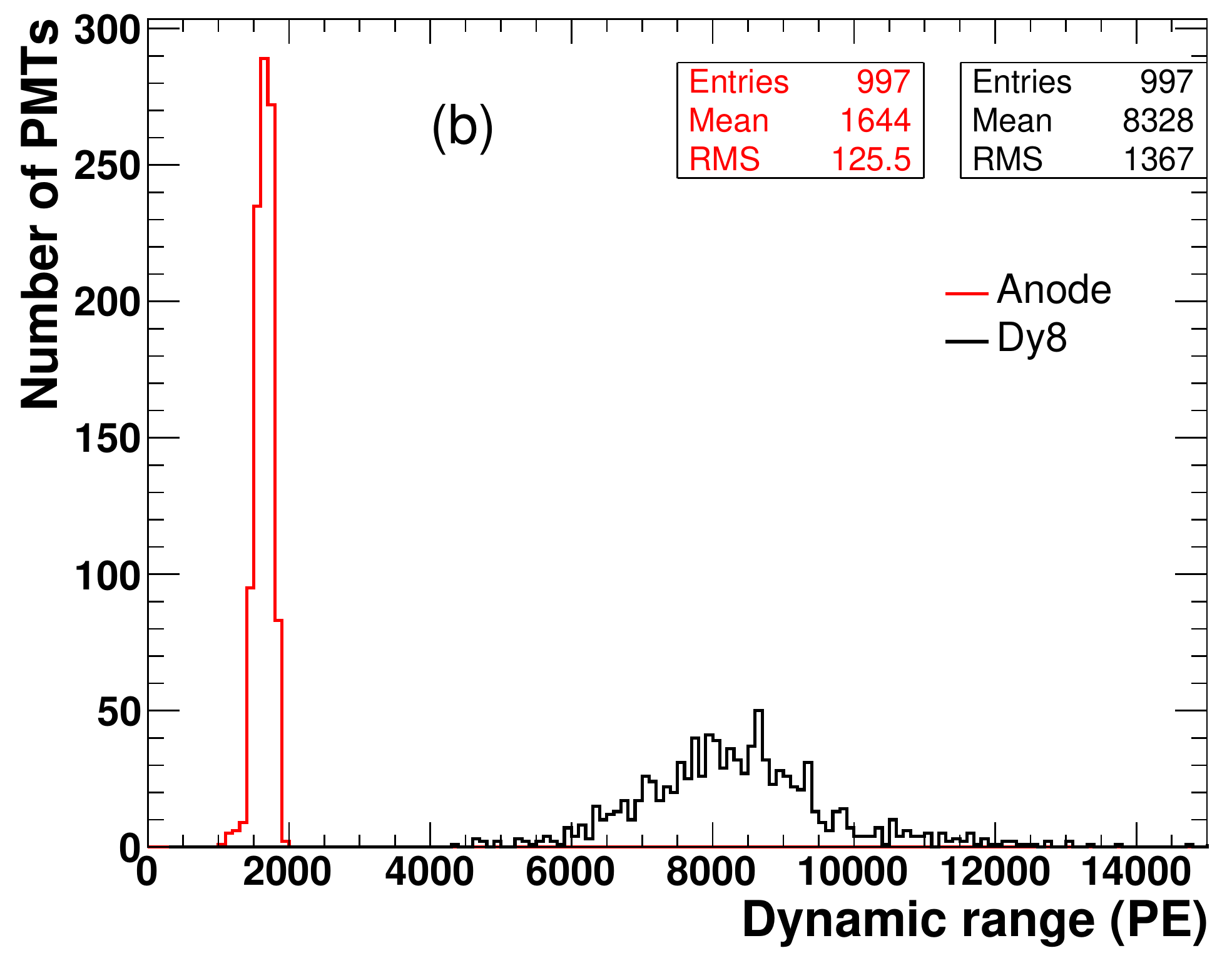}
\vspace*{+0mm}
\caption{(a) Typical non-linearity curves for the anode and the Dy8 as a function of the number of PEs collected by the first dynode. (b) Distributions of dynamic ranges for the anode and the Dy8.} \label{fig:nl}
\end{center}
\end{figure}
}
The A/D is used to convert the charge of Dy8 to the number of PEs collected by the first dynode.
The results show that the anode charge non-linearity is -5\% when the nPE is 1870 and the dynode non-linearity is -5\% when the nPE is 7000.
The dynamic range (nPE) is defined as where the non-linearity exceeds $\pm$5\%.
Fig.~\ref{fig:nl}(b) shows the distributions of dynamic ranges for the anode and the Dy8.  All tested PMTs meet the requirements of dynamic ranges for the anode ($>1000$ PEs) and the Dy8 ($>4000$ PEs).

\section{Systematic uncertainties and test system performance}\label{syserr}
In this program, we measure two kinds of physical quantities: charge and time. The time resolution of the whole system is measured by a fast PMT Hamamatsu R2083 (AA5463). The typical value of TTS for R2083 is 0.37 ns~\cite{r2083}, while the measured value by our system is 0.65 ns. 
This indicates that the upper limit of the time resolution of the system is 0.65 ns.
The mean value of TTS for all tested PMTs is 3.0 ns (Fig.~\ref{fig:tts}(c)). The uncertainty introduced by the time resolution of the test system is less than 0.07 ns ($3.0-\sqrt{3.0^2-0.65^2}$). For charge measurement, the signals are attenuated by the cables and the switches. The gain of the amplifier and the inverter need to be calibrated. The LSB of the QDC may also deviate from the value declared by the manufacturer. To reduce all these systematic uncertainties, the charge output of the generator (Tektronix 3252) is firstly measured by the test system then by an oscilloscope (Agilent MSO9404A). The charge response of the test system is calibrated by the oscilloscope. As mentioned above, the uncertainty in quantum efficiency introduced by assuming all PMTs have the same collection efficiency is less than 5\%.
{
\begin{figure}[htbp]
\begin{center}
\includegraphics[keepaspectratio,width=0.45\textwidth]{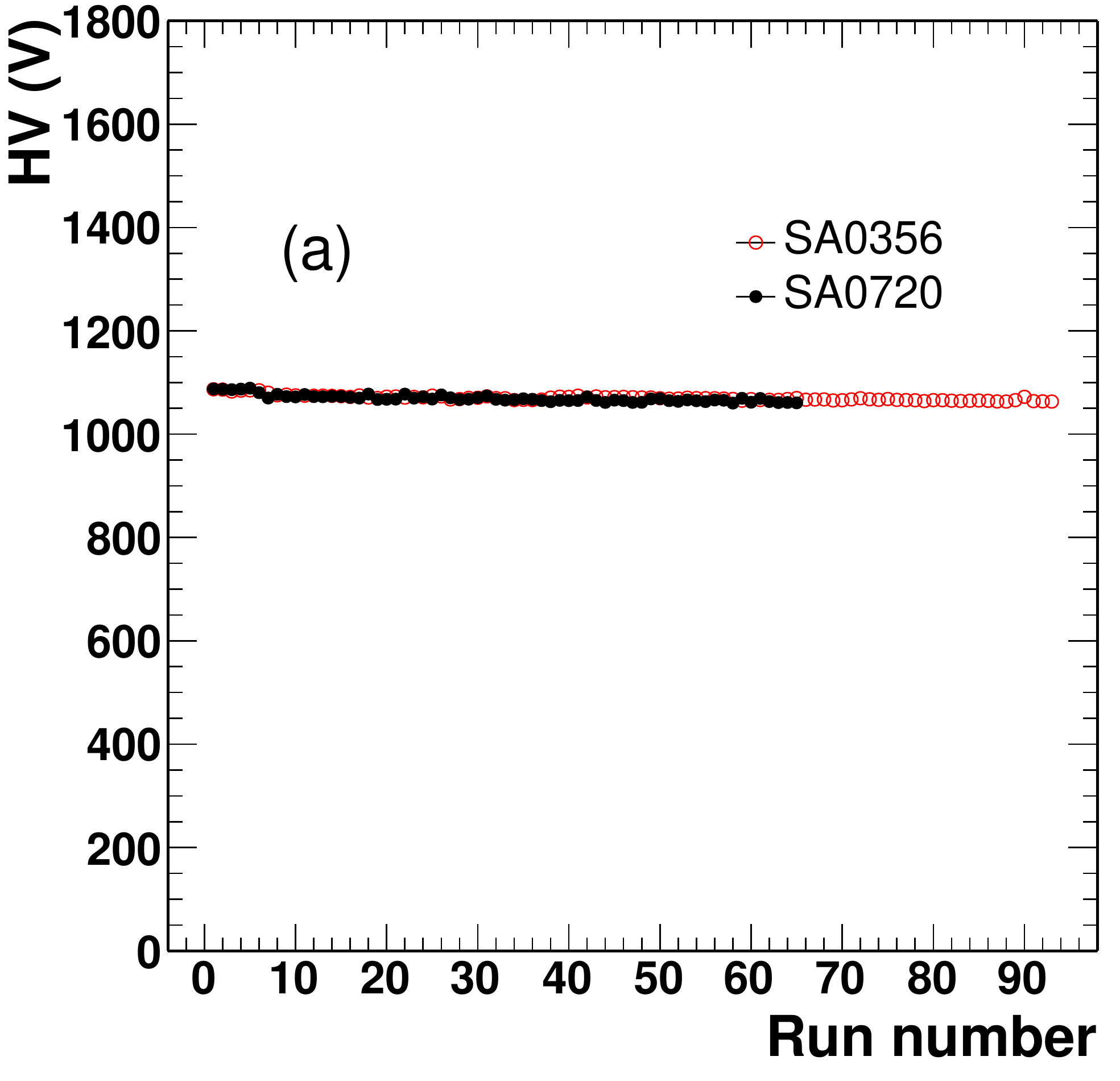}
\includegraphics[keepaspectratio,width=0.45\textwidth]{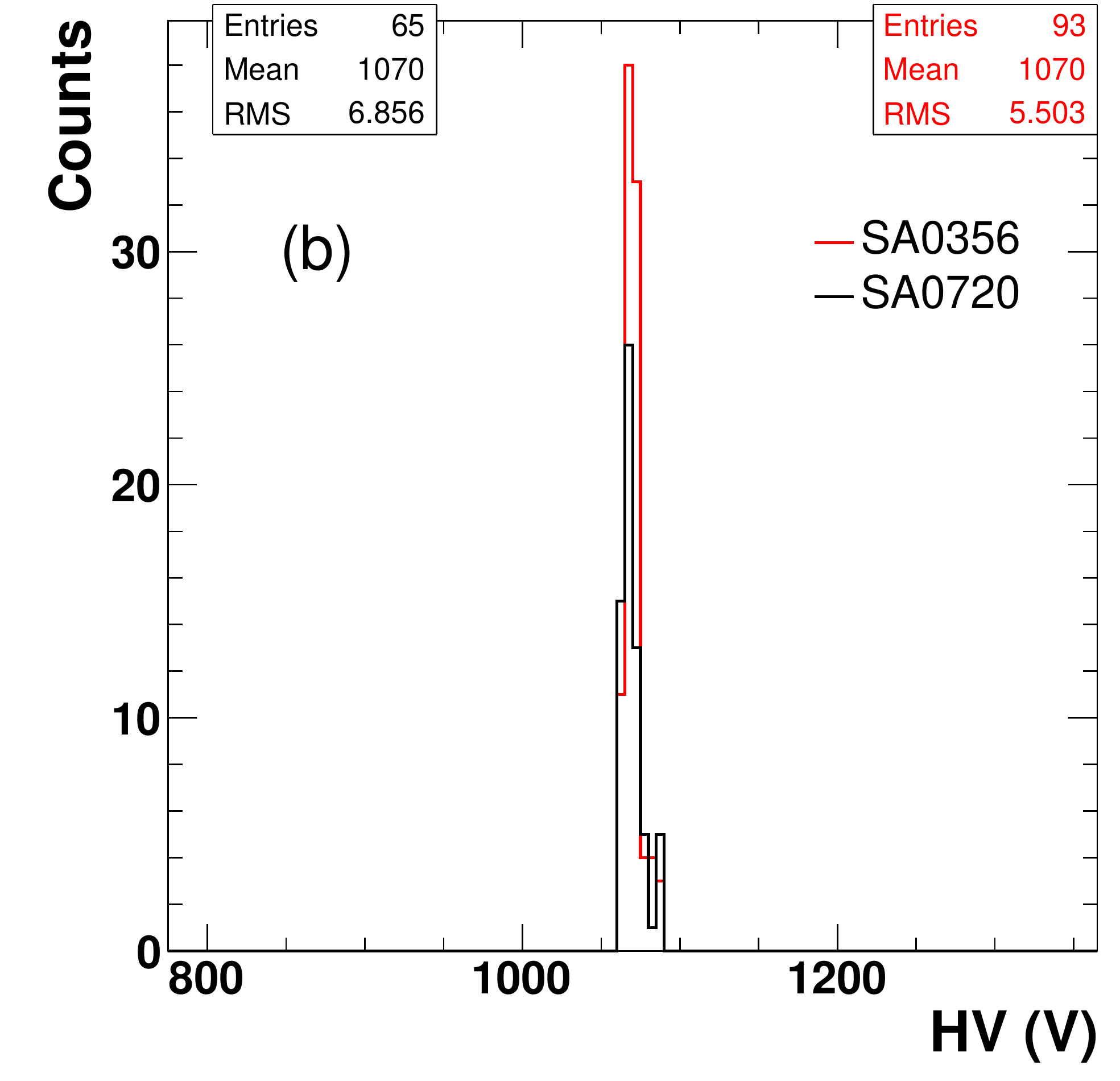}
\vspace*{+0mm}
	\caption{(a) Voltage ($HV_0$) to get a gain of $3\times10^6$ for the two reference PMTs as a function of the run numbers. (b) The distribution of $HV_0$ for the two reference PMTs over nine months} \label{fig:hvref}
\end{center}
\end{figure}
}
{
\begin{figure}[htbp]
\begin{center}
\includegraphics[keepaspectratio,width=0.45\textwidth]{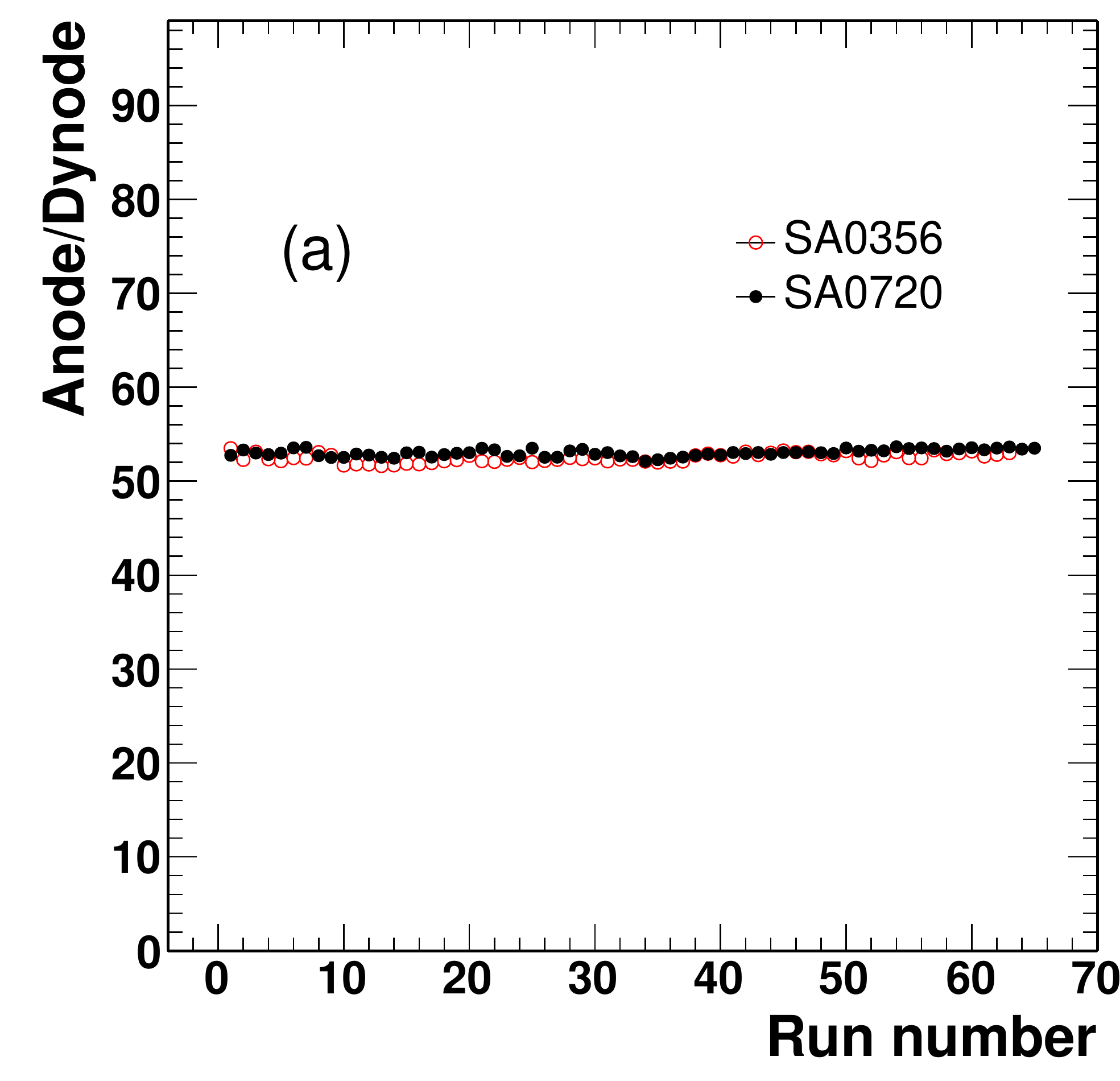}
\includegraphics[keepaspectratio,width=0.45\textwidth]{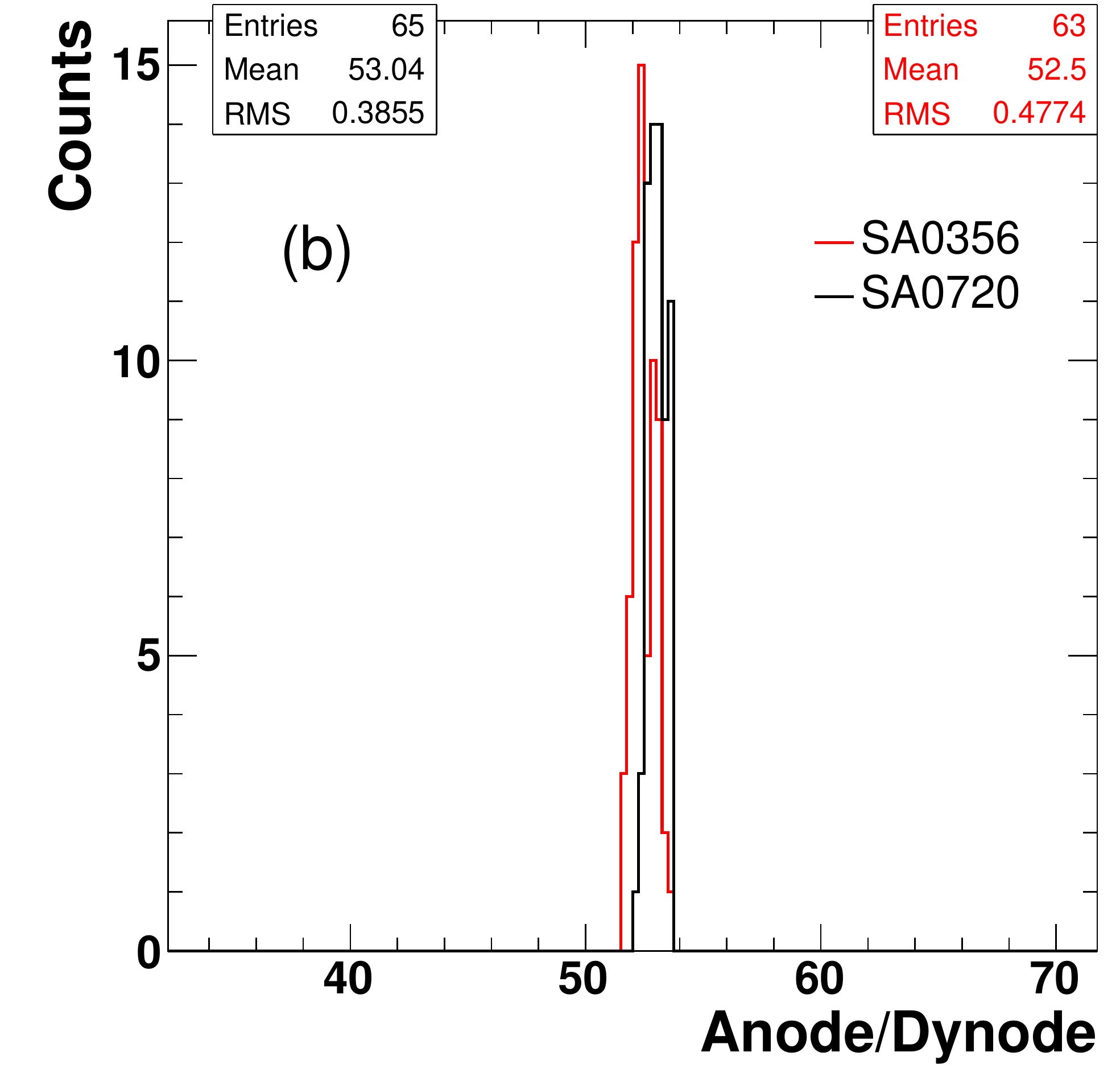}
\vspace*{+0mm}
\caption{(a) The measured A/D for the two reference PMTs as a function of the run numbers. (b) The distribution of A/D for the two reference PMTs over a period of nine months.} \label{fig:adref}
\end{center}
\end{figure}
}
{
\begin{figure}[htbp]
\begin{center}
\includegraphics[keepaspectratio,width=0.45\textwidth]{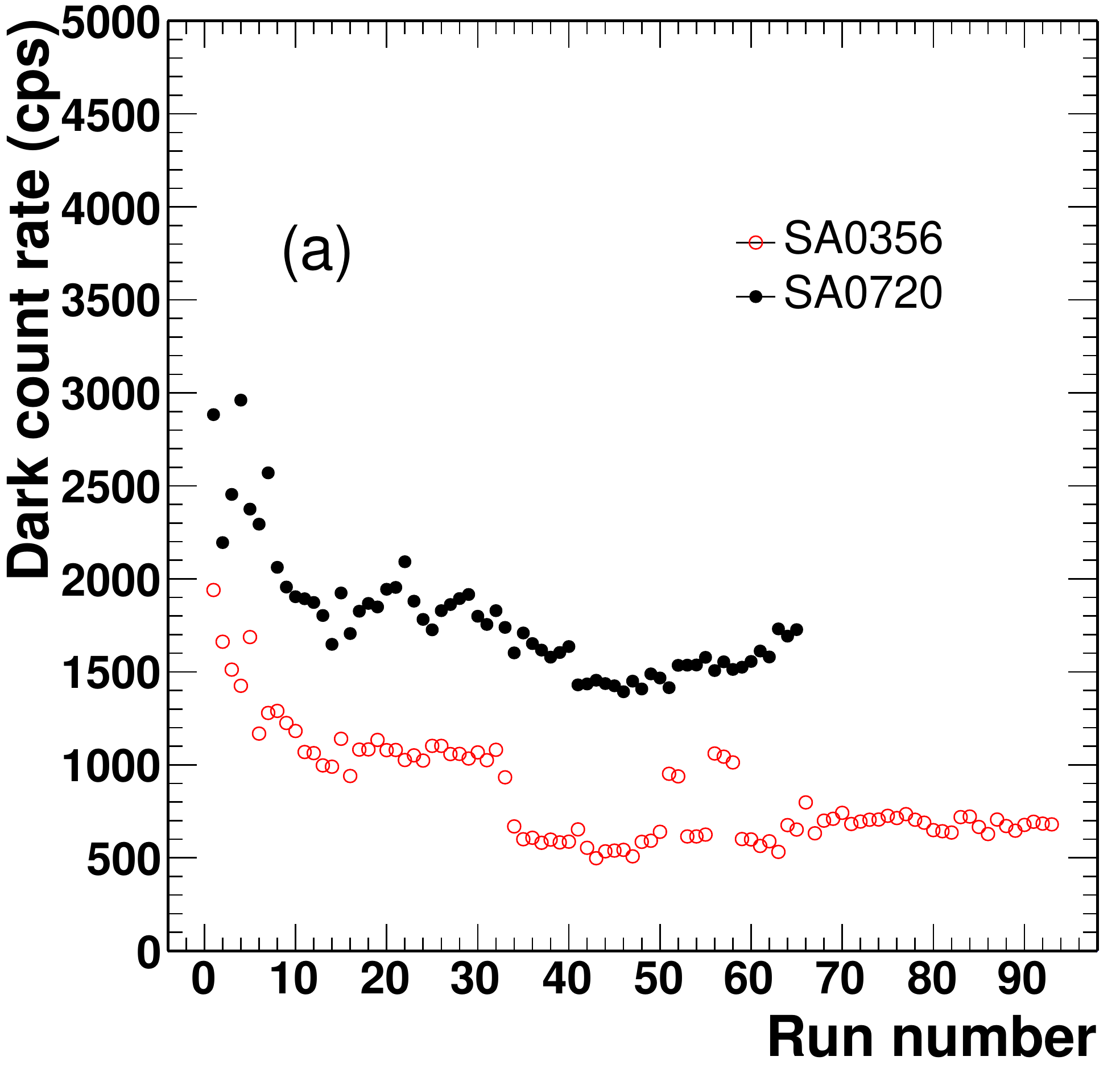}
\includegraphics[keepaspectratio,width=0.45\textwidth]{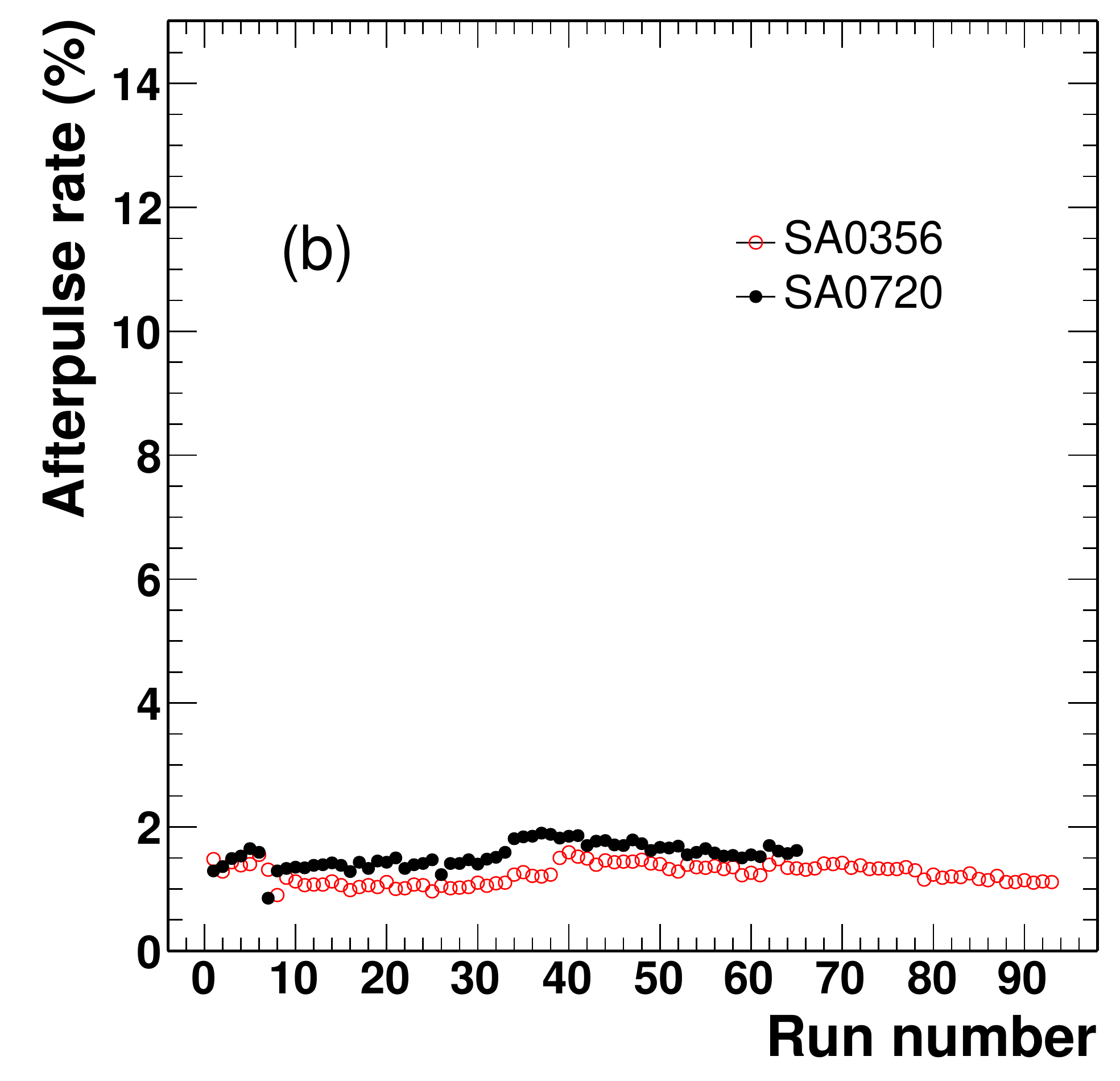}
\includegraphics[keepaspectratio,width=0.45\textwidth]{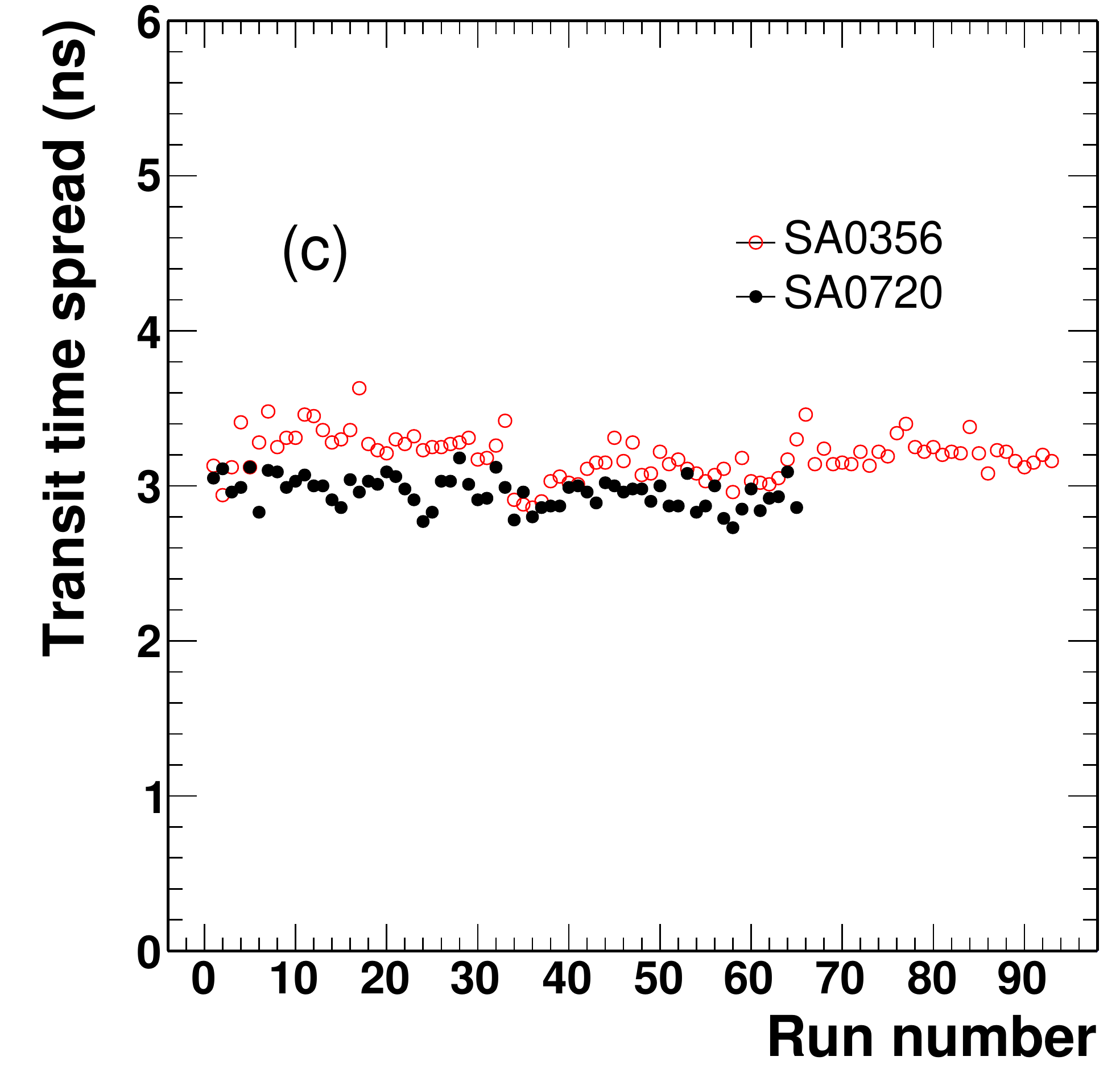}
\includegraphics[keepaspectratio,width=0.45\textwidth]{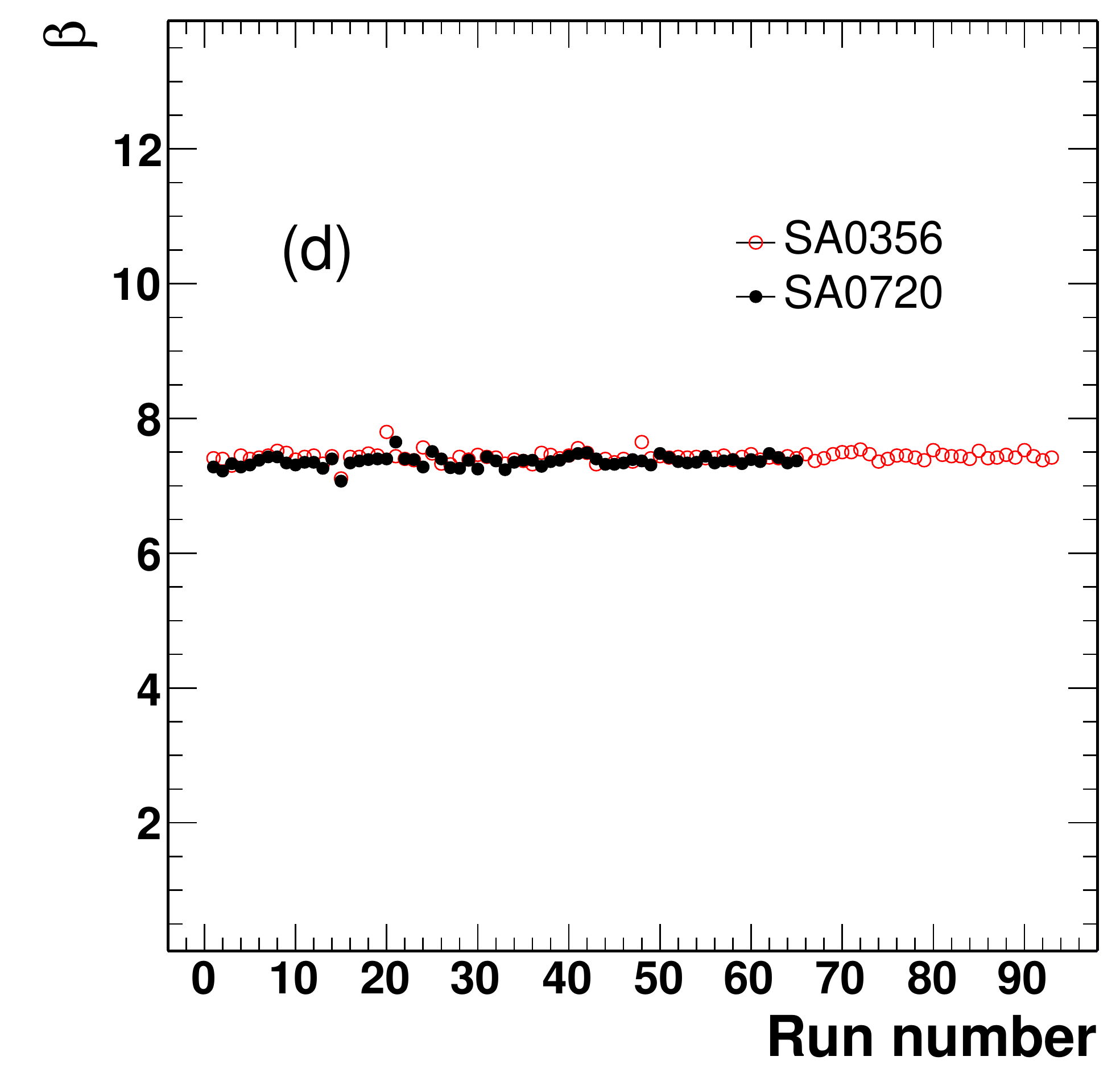}
\vspace*{+0mm}
\caption{The measured dark count rate (a), afterpulse rate (b), transit time spread (c), and amplification voltage coefficient $\beta$ (d) as a function of run numbers for two reference PMTs.} \label{fig:otherpararef}
\end{center}
\end{figure}
}
{
\begin{table*}
\centering
  \begin{tabular}{c|c|c|c}
	  Measurements &SA0356 &SA0720 & LHAASO requirements  \\\hline
	  $HV_0$ (\%) & 0.51 & 0.64 &  $<1$ \\
	  P/V  & 0.14 & 0.15 &  $<0.2$ \\
	  TTS (ns)  & 0.14 & 0.10 &  $<0.2$ \\
	  QE (\%)  & 5.8 & 6.6 &  $<10$ \\
	  Dark count rate (cps) & 348 & 328 &  $<500$ \\
	  Afterpulse rate $(\%)^{\ast}$ & 0.16 & 0.19 &  $<2$ \\
	  A/D (\%)& 0.9 & 0.7 &  $<10$ \\
	  Anode NL $(\%)^{\ast}$ & 0.4 & 0.4 &  $<1.5$ \\
	  Dynode NL $(\%)^{\ast}$ & 0.4 & 0.4 &  $<1.5$ \\
\end{tabular}
	\caption{Resolution of the PMT test system based on the standard deviation of the measurements for the reference PMTs. Values with an asterisk ($^\ast$) denote that the value is the raw spread of the variable which is a measurement in percentage.}
  \label{tab:res}
\end{table*}
}

During the operation, two reference PMTs are constantly measured in each test run.
They are used to monitor the performance of the whole system. In addition, the spread of the measured parameters of the reference PMT is an indication of the resolution of the test system~\cite{Barnhill:2008zz}. Fig.~\ref{fig:hvref}(a) shows the determined $HV_0$ (Eq.~\ref{eq:hv0}) for the two reference PMTs as a function of the consecutive run numbers during nine months. No noticeable drift is observed.
Due to the time constraint, only one reference PMT (SA0356) is put in the test stand for 30 runs. During this time, SA0356 was moved to another DAQ channel where the parameter A/D could not be measured due to damage to QDC2 on this channel. Fig.~\ref{fig:hvref}(b) shows the distribution of $HV_0$ for the two reference PMTs. 
The standard deviations are 5.5 V and 6.9 V for PMT SA0356 and SA0720, which result in resolutions of 0.51\% and 0.64\%, respectively. 
Following the same process, the resolution of the test system is determined for each test and is listed in Table~\ref{tab:res}. 
They all meet the requirements of the LHAASO project. 
Other examples are shown in Fig.~\ref{fig:adref} and Fig.~\ref{fig:otherpararef} to demonstrate the monitoring capabilities of the reference PMTs. 
Fig.~\ref{fig:adref} shows the A/D as a function of run numbers and the distribution of A/D for two reference PMTs.
Fig.~\ref{fig:otherpararef} shows the dark count rate, afterpulse rate, TTS, and amplification voltage coefficient as a function of run numbers for two reference PMTs. The dark count rates show a decreasing trend for the first 40 runs, followed by a period of stability. The afterpulse rate, transit time spread, and amplification voltage coefficient are stable over all time.

\section{Conclusions}\label{conclusions}
The PMT base with a high dynamic range for CR365-02-1 has been designed for the LHAASO-WCDA experiment. The PMT batch test system has been built. The systematic uncertainties and the performance of the test system have been studied. In total, 997 PMTs have been tested with 46 (4.6\%) failing to meet the specifications (see Table~\ref{tab:cut}). Among them, one PMT failed to meet the uniformity of the amplification voltage coefficient $\beta$ ($mean\pm0.5$),
17 PMTs (1.7\%) failed for working voltage $HV_0$ ($mean\pm100$ V),
32 PMTs (3.2\%) failed for anode-to-dynode charge ratio A/D ($mean\pm15\%$), 
and 4 PMTs failed for both $HV_0$ and A/D.

\section*{Acknowledgments}
The research presented in this paper has received strong support from LHAASO collaboration and National Natural Science Foundation of China (No. 11675172, 11775217). We would like to specially thank Prof. Zhen Cao, Huihai He, Zhiguo Yao, Mingjun Chen, Bo Gao, and other members of the LHAASO Collaboration for their valuable support and suggestions.


\bibliography{mybibfile}

\end{document}